\documentclass[twocolumn,Journal]{IEEEtran}
\usepackage[T1]{fontenc}
\usepackage[latin9]{inputenc}
\usepackage{float}
\usepackage{bm}
\usepackage{amsmath}
\usepackage{amsthm}
\usepackage{amssymb}
\usepackage{stackrel}
\usepackage{graphicx}
\usepackage[unicode=true,
 bookmarks=false,
 breaklinks=false,pdfborder={0 0 0},pdfborderstyle={},backref=false,colorlinks=false]
 {hyperref}
\hypersetup{pdftitle={Your Title},
 pdfauthor={Your Name},
 pdfpagelayout=OneColumn, pdfnewwindow=true, pdfstartview=XYZ, plainpages=false}

\makeatletter

\floatstyle{ruled}
\newfloat{algorithm}{tbp}{loa}
\providecommand{\algorithmname}{Algorithm}
\floatname{algorithm}{\protect\algorithmname}

\theoremstyle{plain}
\newtheorem{lem}{\protect\lemmaname}
\theoremstyle{plain}
\newtheorem{thm}{\protect\theoremname}
\theoremstyle{remark}
\newtheorem{rem}{\protect\remarkname}

\usepackage[caption=false,font=footnotesize]{subfig}
\usepackage{algorithmic}
\usepackage{algorithm}

\usepackage{bm}
\usepackage{cite}
\usepackage[english]{babel}
\usepackage[T1]{fontenc}

\IEEEoverridecommandlockouts

\@ifundefined{showcaptionsetup}{}{%
 \PassOptionsToPackage{caption=false}{subfig}}
\usepackage{subfig}
\makeatother

\providecommand{\lemmaname}{Lemma}
\providecommand{\remarkname}{Remark}
\providecommand{\theoremname}{Theorem}

\begin{document}
\title{When to Preprocess? Keeping Information Fresh for Computing Enable
Internet of Things}
\author{Xijun~Wang,~\IEEEmembership{Member,~IEEE,} Minghao~Fang, Chao
Xu,~\IEEEmembership{Member,~IEEE,} Howard H. Yang,~\IEEEmembership{Member,~IEEE,}
Xinghua~Sun,~\IEEEmembership{Member,~IEEE,} Xiang Chen,~\IEEEmembership{Member,~IEEE,}
and Tony Q. S. Quek,~\IEEEmembership{Fellow,~IEEE}\thanks{}\thanks{X. Wang and X. Chen are with School of Electronics and Information
Technology, Sun Yat-sen University, Guangzhou, 510006, China (e-mail:
wangxijun@mail.sysu.edu.cn; chenxiang@mail.sysu.edu.cn). X. Wang is
also with Key Laboratory of Wireless Sensor Network \& Communication,
Shanghai Institute of Microsystem and Information Technology, Chinese
Academy of Sciences, 865 Changning Road, Shanghai 200050 China.}\thanks{M. Fang and X. Sun are with School of Electronics and Communication
Engineering, Sun Yat-sen University, Guangzhou, China (e-mail: fangmh5@mail2.sysu.edu.cn;
sunxinghua@mail.sysu.edu.cn).}\thanks{C. Xu is with School of Information Engineering, Northwest A\&F University,
Yangling, Shaanxi, China (e-mail: cxu@nwafu.edu.cn). C. Xu is also
with Key Laboratory of Agricultural Internet of Things, Ministry of
Agriculture and Rural Affairs, Yangling, Shaanxi, China, and Shaanxi
Key Laboratory of Agricultural Information Perception and Intelligent
Service, Yangling, Shaanxi, China.}\thanks{H. H. Yang is with Zhejiang University/University of Illinois at Urbana-Champaign
Institute, Zhejiang University, Haining 314400, China (e-mail: haoyang@intl.zju.edu.cn)}\thanks{T. Q. S. Quek is with Information System Technology and Design Pillar,
Singapore University of Technology and Design, Singapore 487372 (e-mail:
tonyquek@sutd.edu.sg)}\thanks{Copyright (c) 2021 IEEE. Personal use of this material is permitted.
However, permission to use this material for any other purposes must
be obtained from the IEEE by sending a request to pubs-permissions@ieee.org. }}
\maketitle
\begin{abstract}
Age of information (AoI), a notion that measures the information freshness,
is an essential performance measure for time-critical applications
in Internet of Things (IoT). With the surge of computing resources
at the IoT devices, it is possible to preprocess the information packets
that contain the status update before sending them to the destination
so as to alleviate the transmission burden. However, the additional
time and energy expenditure induced by computing also make the optimal
updating a non-trivial problem. In this paper, we consider a time-critical
IoT system, where the IoT device is capable of preprocessing the status
update before the transmission. Particularly, we aim to jointly design
the preprocessing and transmission so that the weighted sum of the
average AoI of the destination and the energy consumption of the IoT
device is minimized. Due to the heterogeneity in transmission and
computation capacities, the durations of distinct actions of the IoT
device are non-uniform. Therefore, we formulate the status updating
problem as an infinite horizon average cost semi-Markov decision process
(SMDP) and then transform it into a discrete-time Markov decision
process. We demonstrate that the optimal policy is of threshold type
with respect to the AoI. Equipped with this, a structure-aware relative
policy iteration algorithm is proposed to obtain the optimal policy
of the SMDP. Our analysis shows that preprocessing is more beneficial
in regimes of high AoIs, given it can reduce the time required for
updates. We further prove the switching structure of the optimal policy
in a special scenario, where the status updates are transmitted over
a reliable channel, and derive the optimal threshold. Finally, simulation
results demonstrate the efficacy of preprocessing and show that the
proposed policy outperforms two baseline policies. 
\end{abstract}


\section{Introduction}

There is a growing need for real-time status monitoring and controlling
with the overwhelming proliferation of the Internet of Things (IoT),
such as sensor networks, camera networks, and vehicular networks,
to name but a few \cite{palattellaInternetThings5G2016}. Timely updates
of status at the destination are crucial for effective monitoring
and control in these applications \cite{linAverageAgeChanged2020a,xuAoIEnergyConsumption2020}.
As such, we use the metric of age of information (AoI), which is defined
from the receiver\textquoteright s perspective as the time elapsed
since the most recently received status update was generated at the
IoT device \cite{kaulRealtimeStatusHow2012}, to quantify the freshness
of information. In general, minimization of the AoI requires the sampling
frequency, queueing delay, and transmission latency be jointly optimized
at the IoT device, which have been extensively studied in previous
works \cite{sunUpdateWaitHow2017,jiangUnifiedSamplingScheduling2019,zhouJointStatusSampling2019}.

Actually, besides performing simple monitoring tasks, new designed
IoT devices with computing capability is able to conduct more intricate
tasks, such as data compression, feature extraction, and initial classification
\cite{kuangAnalysisComputationIntensiveStatus2020,xuOptimizingInformationFreshness2020}.
Preprocessing the status update at the IoT device can reduce the transmission
time but give rise to an additional preprocessing time. Therefore,
a natural question arises at once: Is it instrumental in reducing
AoI by preprocessing the status updates before the transmission? And
if yes, how to jointly schedule the preprocessing and transmission?
These questions motivate the study of the computing-enable IoT in
this paper.

A recent line of research has exerted substantial efforts in studying
the AoI minimization with computing-enabled IoT devices \cite{kuangAnalysisComputationIntensiveStatus2020,xuOptimizingInformationFreshness2020,zouOptimizingInformationFreshness2019,bastopcuPartialUpdatesLosing2020,bastopcuAgeInformationUpdates2019}.
In \cite{kuangAnalysisComputationIntensiveStatus2020}, the local
computing scheme was analyzed under the zero-wait policy by using
tandem queueing model and was compared with the remote computing scheme
in terms of the average AoI. The tandem queueing model was further
extended in \cite{xuOptimizingInformationFreshness2020}, where the
status updates from multiple sources are preprocessed with different
priorities. The closed-form expression for the average peak AoI was
derived and the effects of the processing rate on the peak AoI was
analyzed. In \cite{zouOptimizingInformationFreshness2019}, both average
AoI and average peak AoI were analyzed for the computing-enable IoT
device with various tandem queueing models, including preemptive and
non-preemptive queueing disciplines. However, these studies are primarily
concerned with the AoI analysis of a computing-enabled IoT system
with a predetermined preprocessing and transmission policy.

The optimal control of the preprocessing at the IoT device has been
studied in \cite{bastopcuPartialUpdatesLosing2020,bastopcuAgeInformationUpdates2019}.
In \cite{bastopcuPartialUpdatesLosing2020}, each status update is
generated with zero-wait policy and preprocessed to regenerate a partial
update. The partial update generation process was optimized to minimize
the average AoI and maintain a desired level of information fidelity.
However, the time consumption of the preprocessing has not been considered.
In \cite{bastopcuAgeInformationUpdates2019}, the processing is used
to improve the quality of the status update at the cost of increasing
the age. Both the waiting time and the processing time were optimized
to find the minimum of the average AoI subject to a desired level
of distortion for each update. Nonetheless, the transmission time
was assumed to be ignorable. 

The status updating problem in a time-critical IoT system is studied
in this paper, where the IoT device is capable of preprocessing the
status updates. In particular, our goal is to control the preprocessing
and transmission procedure jointly  at the IoT device in order to
reduce the weighted sum of the average AoI associated with the destination
and the energy consumed by the IoT device. Under this setup, the IoT
device can stay idle, transmit the status update directly, or preprocess
and transmit the status update. Due to the limited transmission and
computation capacities, each status update takes multiple minislots
to be preprocessed and transmitted. Moreover, because the processing
rate and transmission rate are different in general, the time for
transmitting directly and that for preprocessing-and-transmiting are
unequal. While the model of non-uniform transmission time has also
been investigated in \cite{wangWhenPreemptAge2019,zhouMinimumAgeInformation2020a},
where either the status updates are of different sizes and hence the
durations of the same action may be non-uniform \cite{wangWhenPreemptAge2019},
or the sizes of the status updates are different for different devices
\cite{zhouMinimumAgeInformation2020a}, in this work, it is the duration
of distinct actions that are non-uniform. The key contributions of
this paper are summarized as follows:
\begin{itemize}
\item By accounting for the non-uniform duration of distinct actions, we
formulate the status updating problem as an infinite horizon average
cost semi-Markov decision process (SMDP). In consequence, the Bellman
equation for the uniform time step average cost MDP does not directly
apply. To address this issue, we transform the SMDP to an equivalent
uniform time step MDP. Then, we analyze the structure of the optimal
update policy and put forth a relative policy iteration algorithm
to obtain the optimal update policy based on the structural properties.
We prove that to minimize the long-term average cost, the updating
action with a shorter expected duration should be chosen when the
AoI is large enough to dominate the cost. Therefore, the IoT device
should preprocess the status update before the transmission for large
AoIs, when the preprocessing results in a shorter expected update
duration than direct transmission.
\item The optimal status updating problem is further studied in a special
scenario where the status updates are transmitted over a reliable
channel. Then, we demonstrate that the optimal update policy has a
switch-type structure as to AoI in two cases. In the first case, the
action of being idle is excluded in the optimal policy, while in the
second case the action with lower energy efficiency is excluded in
the optimal policy. The optimal thresholds are further derived in
both cases.
\item We evaluate the performance of the optimal update policy and compare
it with two zero-wait policies by conducting extensive simulations.
The results demonstrate that the optimal update policy can effectively
schedule the preprocessing and transmission and walk a fine line between
the AoI and the energy consumption. 
\end{itemize}

The rest of the paper is organized as follows: Section II presents
the system model. In Section III, we provide the SMDP formulation
of the problem and propose the structure-aware relative policy iteration
algorithm. In Section IV, we study the structure property of the optimal
policy in a special scenario. In Section V, the simulation results
are discussed, followed by the conclusion in Section VI.

\section{System Model \label{sec:System-Overview}}

\begin{figure}[tp]
\centering

\includegraphics[width=0.5\textwidth]{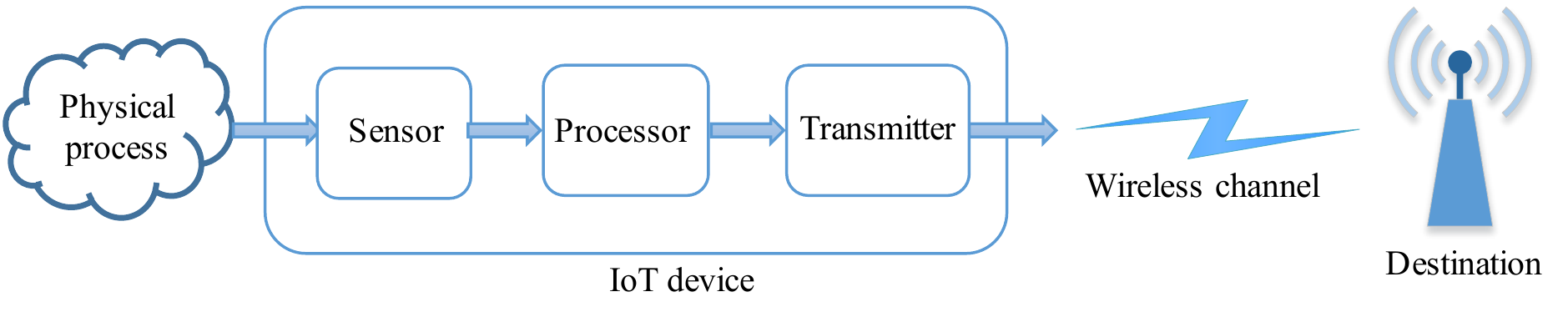}\caption{\label{fig:SystemModel}An illustration of the IoT status monitoring
system.}
\end{figure}

As illustrated in Fig. \ref{fig:SystemModel}, we consider a time-critical
IoT status updating system with a single IoT device and a destination.\footnote{Although we consider only one device, the result in this paper can
be extended to the IoT system with multiple devices by formulating
the status updating problem as a restless multi-armed bandit (RMAB)
problem. To solve the RMAB problem, Whittle\textquoteright s index
policy can be employed, where we decouple the problem with multiple
devices into multiple sub-problems. There is only a single source-destination
pair in each sub-problem, which is exactly the model we considered
in this work.} The IoT device is composed of a sensor which is capable of tracking
the status of the underlying physical process, a processor which is
capable of preprocessing the status update, and a transmitter which
can deliver the status update over a wireless channel to the destination.
The model with a single source-destination pair is simple but sufficient
enough to investigate a wide range of applications. We assume that
the IoT device adheres to the generate-at-will policy, which implies
that a fresh status update is generated anytime an update decision
is made.

A time-slotted system is considered, where time is divided into minislots
with equal duration of $\tau$ (in seconds). In this system, a status
update with $T_{u}$ packets is generated at the beginning of a minislot
and at most one packet can be transmitted in one minislot. As such,
the total duration for transmitting a single status update is $T_{u}$
minislots. The preprocessing at the IoT device could be data compression,
feature extraction, or initial classification. In this work, we consider
the preprocessing in general practice. Specifically, we characterize
the preprocessing operation with three parameters, namely, the size
of the status update before preprocessing $T_{u}$, the size of the
status update after preprocessing $T_{u}'$, and the number of CPU
cycles per bit required to complete this operation $v$.\footnote{Here, we would like to take the data compression as an example to
explain the relationship between these parameters. For data compression,
$T_{u}$ and $T_{u}'$ are related to each other with a data compression
ratio $\beta$, i.e., $T_{u}'=\beta T_{u}$. Moreover, to perform
the compression operation with the ratio $\beta$, the number of CPU
cycles required to compress one bit of the input data is $v$. } Let $l$ denote the number of bits per packet. Since the number
of bit of the status update before processing is $T_{u}l$, the number
of minislots required for preprocessing one status update is then
given by 
\begin{equation}
T_{p}=\left\lceil \frac{T_{u}l\upsilon}{f\tau}\right\rceil ,
\end{equation}
where $f$ (in Hz) is the CPU frequency of the processor. We assume
that the destination (e.g., a base station or an access point) has
a more powerful computing capability. Therefore, the processing time
at the destination is negligible compared to the processing time at
the IoT device or the transmission time. 

\begin{figure}[tp]
\centering

\includegraphics[width=0.5\textwidth]{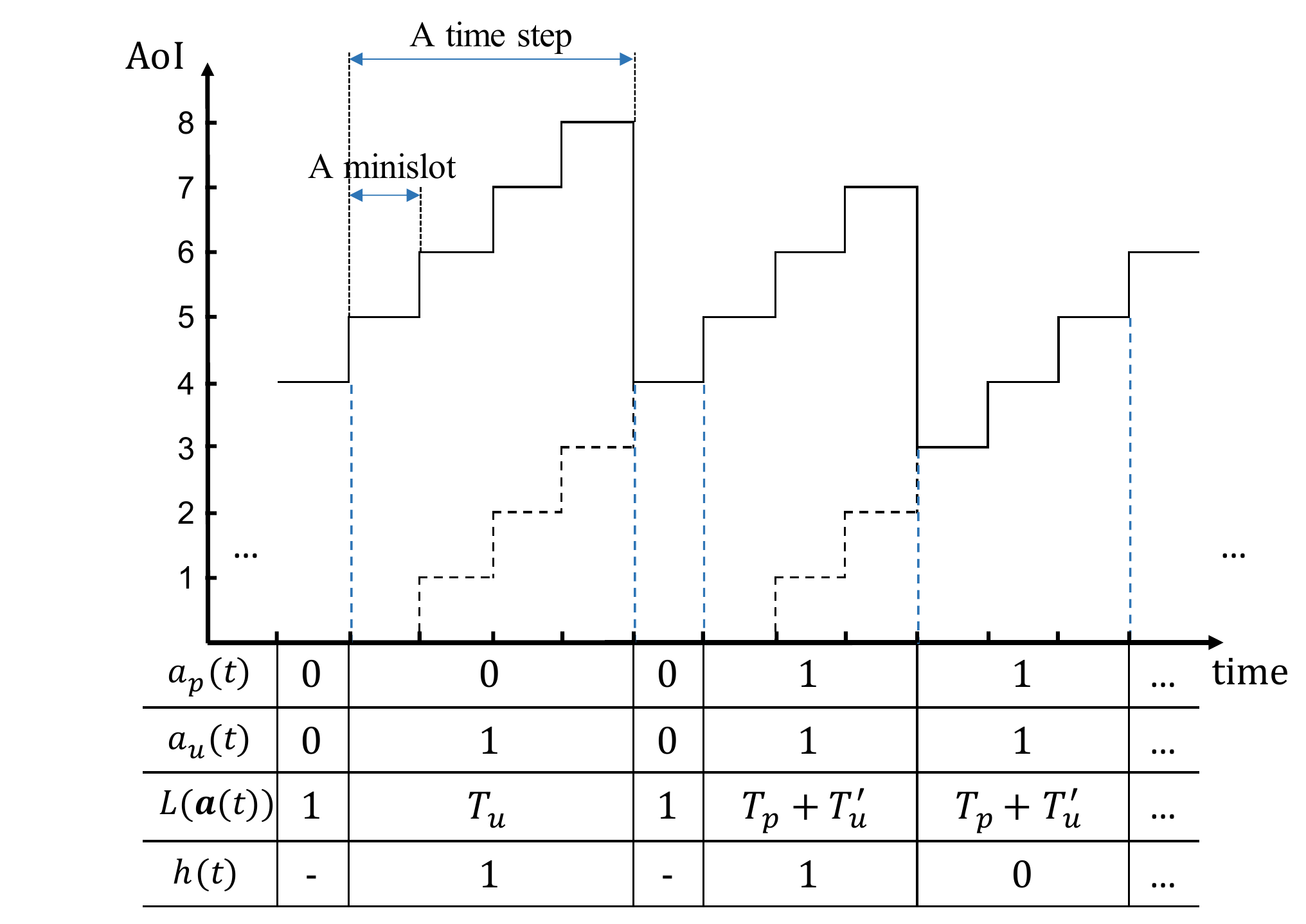}\caption{\label{fig:AoIEvolution}An illustration of the evolution of the AoI,
where $T_{u}=4$, $T_{u}'=2$, $T_{p}=1$.}
\end{figure}

We refer to a decision epoch of the IoT device as a time step, as
illustrated in Fig. \ref{fig:AoIEvolution}. In each time step, the
IoT device must determine whether to sample and transmit an update
directly or preprocess the update before the transmission. Let $a_{p}(t)\in\{0,1\}$
denote the computing action at time step $t$, where $a_{p}(t)=1$
indicates that the device preprocesses the status update, and $a_{p}(t)=0$,
otherwise. Let $a_{u}(t)\in\{0,1\}$ denote the updating action at
time step $t$, where $a_{u}(t)=1$ indicates that the device samples
and transmits the status update to the destination and $a_{u}(t)=0$,
otherwise. Let $\boldsymbol{a}(t)\triangleq(a_{p}(t),a_{u}(t))\in\mathcal{A\triangleq}\left\{ (0,0),(0,1),(1,1)\right\} $
denote the device's control action vector  at time step $t$, where
$\mathcal{A}$ is the feasible action space. In particular, if $\boldsymbol{a}(t)=(0,0)$,
the device will stay idle in one minislot. If $\boldsymbol{a}(t)=(0,1)$,
the device will sample and transmit the update directly without preprocessing.
If $\boldsymbol{a}(t)=(1,1)$, the device will first preprocess the
status update after sampling and then transmit it to the destination.
Notably, the action vector $(1,0)$ is not feasible because this action
incurs energy consumption but does not provide the destination with
a fresh status update. 

It is important to emphasize that the duration of a time step is not
uniform.  Specifically, let $L(\boldsymbol{a}(t))$ denote the number
of minislots in time step $t$ with action $\bm{a}(t)$ being taken,
we can then express $L(\boldsymbol{a}(t))$ as follows 
\begin{equation}
L(\boldsymbol{a}(t))=\begin{cases}
1, & \text{if }\boldsymbol{a}(t)=(0,0),\\
T_{u}, & \text{if }\boldsymbol{a}(t)=(0,1),\\
T_{p}+T_{u}', & \text{if }\boldsymbol{a}(t)=(1,1).
\end{cases}\label{eq:Duration}
\end{equation}
We further denote by $L_{u}(\boldsymbol{a}(t))$ the transmission
time corresponding to action $\bm{a}(t)$, which is given as follows
\begin{equation}
L_{u}(\boldsymbol{a}(t))=\begin{cases}
0, & \text{if }\boldsymbol{a}(t)=(0,0),\\
T_{u}, & \text{if }\boldsymbol{a}(t)=(0,1),\\
T_{u}', & \text{if }\boldsymbol{a}(t)=(1,1).
\end{cases}\label{eq:TransDuration}
\end{equation}

Let $C_{p}$ denote the computation energy consumption per minislot
when $a_{p}(t)=1$ and $C_{u}$ denote the communication energy consumption
per minislot when $a_{u}(t)=1$. In particular, the computation energy
consumption per minislot is given by
\begin{equation}
C_{p}=\kappa\tau f^{3},
\end{equation}
where $\kappa$ is the effective switched capacitance depending on
the chip architecture. By assuming a constant transmission power $P$
of the IoT device, the communication energy consumption per minislot
is $C_{u}=P\tau$. Then, the total energy consumption associated with
action $\bm{a}(t)$ at time step $t$ is given by 
\begin{align}
C(\bm{a}(t)) & =\begin{cases}
0, & \text{if }\boldsymbol{a}(t)=(0,0),\\
T_{u}C_{u}, & \text{if }\boldsymbol{a}(t)=(0,1),\\
T_{p}C_{p}+T_{u}'C_{u}, & \text{if }\boldsymbol{a}(t)=(1,1).
\end{cases}
\end{align}

It is assumed that channel fading is constant in each minislot but
varies independently across them. The channel state information is
also assumed to be available only at the destination and the IoT device
transmits an update at a fixed rate. We use a memoryless Bernoulli
process $h(t,i)\in\{0,1\}$ to characterize the transmission failure
because of outage, where $h(t,i)=1$ indicates that the packet is
transmitted successfully at the $i$-th minislot of time step $t$,
and $h(t,i)=0$, otherwise. The transmission success probability of
a packet is defined as 
\begin{equation}
p_{s}=\Pr\{h(t,i)=1\}=\Pr\left\{ B\log\left(1+\frac{\gamma P}{\sigma^{2}}\right)\geq\frac{l}{\tau}\right\} ,
\end{equation}
where $B$ is the channel bandwidth, $\gamma$ is the channel gain
between the IoT device and the destination, and $\sigma^{2}$ is the
noise power. We assume that the status update can be successfully
recovered at the destination if all the packets are transmitted successfully
during one time step. We denote by $h(t)\in\{0,1\}$ the transmission
status of an update at time step $t$, i.e., $h(t)=\prod_{i=L(\boldsymbol{a}(t))-L_{u}(\boldsymbol{a}(t))+1}^{L(\boldsymbol{a}(t))}h(t,i)$,
where $h(t)=1$ indicates that the update is transmitted successfully,
and $h(t)=0$, otherwise. Thus, the transmission success probability
of an update is given by $\Pr\{h(t)=1\}=p_{s}^{L_{u}(\boldsymbol{a}(t))}$
and the transmission failure probability of an update is given by
$\Pr\{h(t)=0\}=1-p_{s}^{L_{u}(\boldsymbol{a}(t))}$. We assume that
there exists an instantaneous error-free single-bit ACK/NACK feedback
from the destination to the IoT device. After a status update arrives
at the destination, an ACK signal (a NACK signal) is sent in case
of a successful reception (a failure).  

The freshness of the status update is measured via AoI, which is defined
as the time elapsed since the generation of the most recently received
status update. Formally, let $U(t)$ denote the time step at which
the most up-to-date status update successfully received by the destination
was generated. Then, the AoI at the $i$-th minislot of time step
$t$ can be defined as 
\begin{align}
\delta(t,i) & =\stackrel[n=U(t)]{t-1}{\sum}L(\boldsymbol{a}(n))+i-1,
\end{align}
where the first term represents the number of minislots in the previous
time steps since $U(t)$ and $i-1$ is the number of minislots in
the current time step. For simplicity, we represent the AoI at the
beginning of time step $t$ as $\delta(t)$, i.e., $\delta(t)=\delta(t,1)=\sum_{n=U(t)}^{t-1}L(\boldsymbol{a}(n))$. 

Since it is pointless to receive a status update with a very large
age for time-critical IoT application, we let $\hat{\delta}$ be
the upper limit of the AoI, which is assumed to be finite but arbitrarily
large \cite{zhouJointStatusSampling2019}. Then, we present the dynamics
of the AoI as follows
\begin{align}
 & \delta(t+1)=\nonumber \\
 & \begin{cases}
\min(\delta(t)+1,\hat{\delta}), & \text{if }\boldsymbol{a}(t)=(0,0),\\
\min(T_{u},\hat{\delta}), & \text{if }\boldsymbol{a}(t)=(0,1)\text{ and }h(t)=1,\\
\min(\delta(t)+T_{u},\hat{\delta}), & \text{if }\boldsymbol{a}(t)=(0,1)\text{ and }h(t)=0,\\
\min(T_{p}+T_{u}',\hat{\delta}), & \text{if }\boldsymbol{a}(t)=(1,1)\text{ and }h(t)=1,\\
\min(\delta(t)+T_{p}+T_{u}',\hat{\delta}), & \text{if }\boldsymbol{a}(t)=(1,1)\text{ and }h(t)=0.
\end{cases}\label{eq:Dynamic}
\end{align}
We also illustrate the AoI evolution process in Fig. \ref{fig:AoIEvolution}. 

\section{Optimal Update Algorithm }

\subsection{SMDP Formulation}

Since the duration of each time step depends on the action taken in
that time step, the time interval between two sequential actions is
inconstant. Therefore, the optimal updating problem belongs to the
class of SMDP. An SMDP can be defined as a tuple $(\mathcal{S},\mathcal{A},t^{+},\Pr(\cdot|\cdot),R)$,
where $\mathcal{S}$ is the state space, $\mathcal{A}$ is the action
space, $t^{+}$ is the decision epoch, $\Pr(\cdot|\cdot)$ is the
transition probability, and $r$ is the cost function. In particular,
at the beginning of the time step $t$, the agent observes the system
state $s(t)$ and chooses an action $\bm{a}(t)$. As a consequence,
the system remains at $s(t)$ until the next decision epoch. Then,
the system state transitions to $s(t+1)$ and the agent receives a
cost $R(t)$. We note that this is different from MDP, where the transition
time is fixed and independent of the actions. In the following, we
formally define the state, action, transition probability, and cost
function of the SMDP. 

\subsubsection{State }

The state of the SMDP at time step $t$ $s(t)$ is defined to be the
AoI at the beginning of that time step, i.e., $s(t)=\delta(t)$. Since
we limit the maximum value of the AoI, the state space is expressed
as $\mathcal{S}\triangleq\{1,2,\cdots,\hat{\delta}\}$. 

\subsubsection{Action}

The action at time step $t$ is $\boldsymbol{a}(t)$ and the action
space is $\mathcal{A\triangleq}\left\{ (0,0),(0,1),(1,1)\right\} $. 

\subsubsection{Decision Epoch}

A decision is making at the beginning of a time step. The time interval
between two adjacent decision epochs is $L(\bm{a}(t))$, which depends
on the action taking in time step $t$. 

\subsubsection{Transition Probability}

We denote by $\Pr(s(t+1)\mid s(t),\boldsymbol{a}(t))$ the transition
probability that a state transits from $s(t)$ to $s(t+1)$ with action
$\boldsymbol{a}(t)$. According to the AoI evolution dynamic in (\ref{eq:Dynamic}),
the transition probability can be given as follows
\begin{gather}
\begin{cases}
\Pr\left(\min(\delta(t)+1,\hat{\delta})\mid\delta(t),(0,0)\right)=1,\\
\Pr\left(\min(T_{u},\hat{\delta})\mid\delta(t),(0,1)\right)=p_{s}^{T_{u}},\\
\Pr\left(\min(\delta(t)+T_{u},\hat{\delta})\mid\delta(t),(0,1)\right)=1-p_{s}^{T_{u}},\\
\Pr\left(\min(T_{p}+T_{u}',\hat{\delta})\mid\delta(t),(1,1)\right)=p_{s}^{T_{u}'},\\
\Pr\left(\min(\delta(t)+T_{p}+T_{u}',\hat{\delta})\mid\delta(t),(1,1)\right)=1-p_{s}^{T_{u}'}.
\end{cases}
\end{gather}

\subsubsection{Cost}

We aim to minimize the weighted sum of the average AoI associated
with the destination and the energy consumed by the IoT device. As
such, we define the cost at a time step as the weighted sum of the
AoI and the energy consumption. Specifically, the cost at time step
$t$ is represented as 
\begin{align}
 & R(\delta(t),\boldsymbol{a}(t))\nonumber \\
= & \stackrel[i=1]{L(\boldsymbol{a}(t))}{\sum}\delta(t,i)+\omega C(\bm{a}(t))\nonumber \\
= & \stackrel[i=1]{L(\boldsymbol{a}(t))}{\sum}(\delta(t)+i-1)+\omega C(\bm{a}(t))\nonumber \\
= & \frac{1}{2}(2\delta(t)+L(\boldsymbol{a}(t))-1)L(\boldsymbol{a}(t))+\omega C(\bm{a}(t)),\label{eq:total-cost}
\end{align}
where $\omega$ is the weighting factor.

Our goal is to find an update policy $\pi=(\boldsymbol{a}(1),\boldsymbol{a}(2),\ldots)$
that reduces the average cost to the lowest possible level. Under
a set of stationary deterministic policy $\Pi$ and a given initial
system state $s(1)$, the objective can be formulated as follows:
\begin{align}
\min_{\pi\in\Pi}\limsup_{T\rightarrow\infty}\frac{\mathbb{E}\left[\stackrel[t=1]{T}{\sum}R(\delta(t),\boldsymbol{a}(t))\mid s(1)\right]}{\mathbb{E}\left[\stackrel[t=1]{T}{\sum}L(\boldsymbol{a}(t))\right]}.\label{eq:Problem}
\end{align}
Since the duration of the time step is not uniform, the average cost
in (\ref{eq:Problem}) is defined as the limit of the expected total
cost over a finite number of time steps divided by the expected cumulative
time of these time steps \cite{sunUpdateWaitHow2017}. In this work,
we restrict our attention to stationary unichain policy, under which
the Markov chain is composed of a single recurrent class and a set
of transient states. Thus, the average cost is independent on the
initial state and the Markov chain has a unique stationary distribution
\cite{putermanMarkovDecisionProcesses2005}.

To solve this problem, we transform the SMDP into an equivalent discrete
time MDP using uniformization \cite{tijmsSemiMarkovDecisionProcesses2004,putermanMarkovDecisionProcesses2005}.
Let $\mathcal{\bar{S}}$ and $\mathcal{\bar{A}}$ denote the state
space and action space of the transformed MDP. They are the same as
those in the original SMDP, i.e., $\mathcal{\bar{S}}=\mathcal{S}$
and $\mathcal{\bar{A}}=\mathcal{A}$. For any $s\in\bar{\mathcal{S}}$
and $\boldsymbol{a}\in\bar{\mathcal{A}}$, the cost in the MDP is
given by
\begin{equation}
\bar{R}(s,\boldsymbol{a})=\frac{R(s,\boldsymbol{a})}{L(\boldsymbol{a})}=s+\frac{1}{2}(L(\boldsymbol{a})-1)+\omega\frac{C(\bm{a})}{L(\boldsymbol{a})},
\end{equation}
and the transition probability is given by
\begin{equation}
\bar{p}(s'\mid s,\boldsymbol{a})=\begin{cases}
\frac{\epsilon}{L(\boldsymbol{a})}p(s'\mid s,\boldsymbol{a}), & s'\neq s,\\
1-\frac{\epsilon}{L(\boldsymbol{a})}, & s'=s,
\end{cases}
\end{equation}
where $\epsilon$ is chosen in $\Big(0,\min\limits _{\boldsymbol{a}}L(\boldsymbol{a})\Big]$. 

Then, by solving the Bellman equation in (\ref{eq:bellman-equ}),
one can obtain the optimal policy $\pi^{*}$ of the original SMDP
that minimizes the average cost. According to \cite{tijmsSemiMarkovDecisionProcesses2004},
we have
\begin{equation}
\theta+V(s)=\min\limits _{\boldsymbol{a}\in\mathcal{A}}\bigg\{\bar{R}(s,\boldsymbol{a})+\sum\limits _{s'\in\mathcal{S}}\bar{p}(s'\mid s,\boldsymbol{a})V(s')\bigg\},\forall s\in\mathcal{S},\label{eq:bellman-equ}
\end{equation}
where $\theta$ is the optimal value to (\ref{eq:Problem}) for all
initial state and $V(s)$ is the value function for the discrete-time
MDP. Then, the optimal policy can be given by
\begin{equation}
\pi^{*}(s)=\arg\min\limits _{\boldsymbol{a}\in\mathcal{A}}\bigg\{\bar{R}(s,\boldsymbol{a})+\sum\limits _{s'\in\mathcal{S}}\bar{p}(s'\mid s,\boldsymbol{a})V(s')\bigg\}\label{eq:opt-policy}
\end{equation}
for any $s\in\mathcal{S}$. Theoretically, we can obtain the optimal
policy $\pi^{*}$ via (\ref{eq:opt-policy}). However, the value function
$V(\cdot)$ does not have closed-form solution in general, which makes
this problem challenging. Although numerical algorithms such as value
iteration and policy iteration can solve this problem, they incur
high computational complexity and do not provide many design insights.
For a better understanding of the system, we will investigate the
structural properties of the optimal update policy in the next subsection.

\subsection{Structural Analysis and Algorithm Design}

In this subsection, we first show that the structure of the optimal
policy is of threshold-type with respect to the AoI. Then, we propose
a relative policy iteration algorithm based on the threshold structure
to obtain the optimal policy $\pi^{*}$ for (\ref{eq:Problem}).

To begin with, we show some key properties of the value function $V(s)$
in the following lemmas. 
\begin{lem}
\label{lem:lem1}The value function $V(s)$ is non-decreasing with
$s$.
\end{lem}
\begin{IEEEproof}
See Appendix \ref{subsec:Proof-of-Lem1}.
\end{IEEEproof}
\begin{lem}
\label{lem:lem2}The value function $V(s)$ is concave in $s$.
\end{lem}
\begin{IEEEproof}
See Appendix \ref{subsec:Proof-of-Lem2}.
\end{IEEEproof}
Since $V(s)$ is a concave function, its slope is non-increasing.
We drive the lower bound of the slope of $V(s)$ in the following
lemma. Before that, we define an auxiliary variable $\bm{a}_{f}$,
which is given by
\begin{equation}
\boldsymbol{a}_{f}=\begin{cases}
(0,1), & \frac{T_{u}}{p_{s}^{T_{u}}}\leq\frac{T_{p}+T_{u}'}{p_{s}^{T_{u}'}},\\
(1,1), & \frac{T_{u}}{p_{s}^{T_{u}}}\geq\frac{T_{p}+T_{u}'}{p_{s}^{T_{u}'}}.
\end{cases}
\end{equation}

\begin{figure}[tp]
\centering

\includegraphics[scale=0.55]{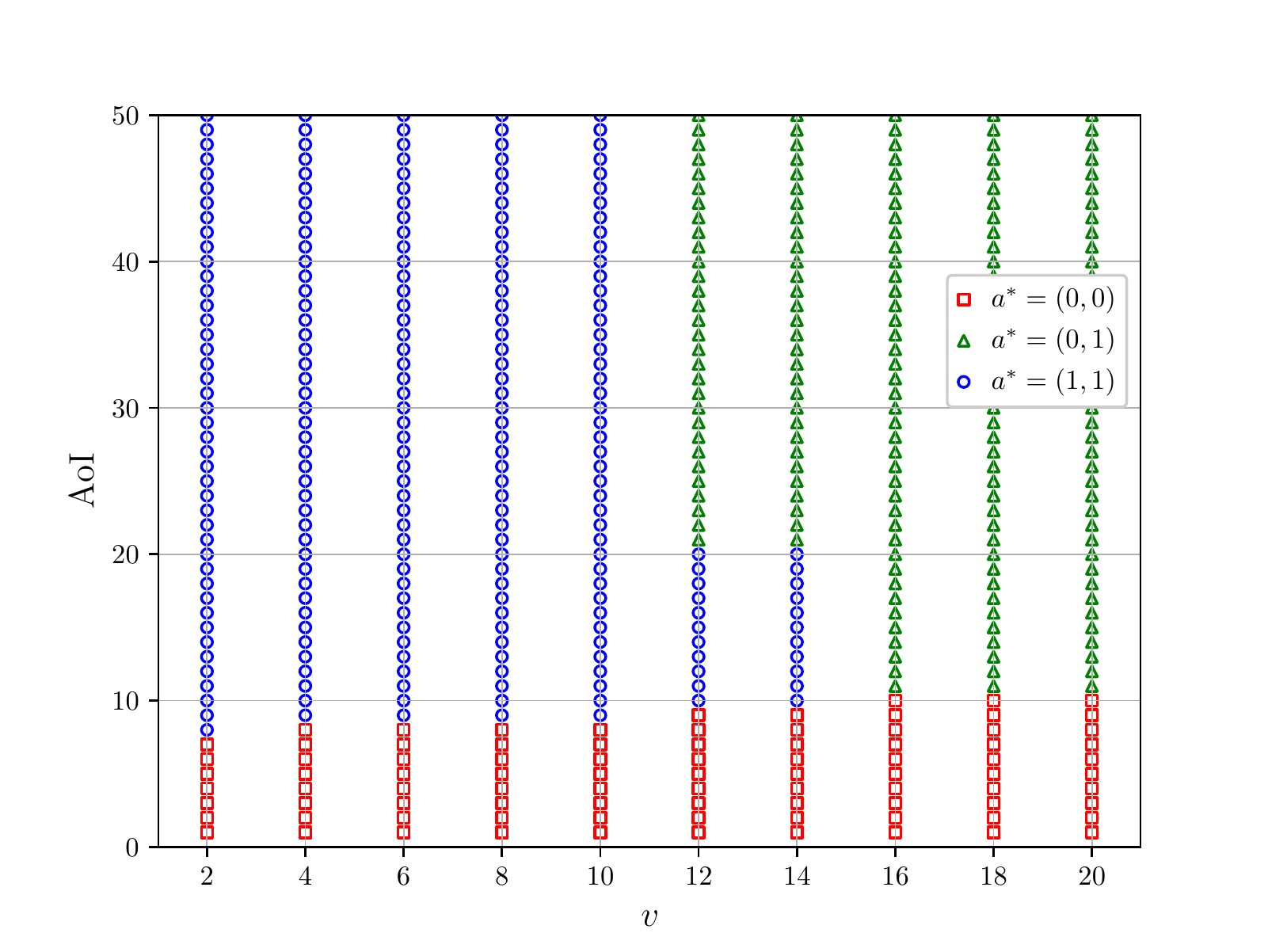}

\caption{\label{fig:structure-general}Structure of the optimal policy for
different values of $v$ ($T_{u}=4$, $T_{u}'=2$, $l=3$, $f=35$,
$\tau=1$, $\kappa=0.00005$, $P=6$, $\omega=2$, $p_{s}=0.8$). }
\end{figure}

\begin{lem}
\label{lem:lem3}For any $s_{1},s_{2}\in\mathcal{S}$, such that $s_{1}\le s_{2}$,
$V(s_{2})-V(s_{1})\geq\frac{L(\boldsymbol{a}_{f})}{\epsilon p_{s}^{L_{u}(\boldsymbol{a}_{f})}}(s_{2}-s_{1})$.
\end{lem}
\begin{IEEEproof}
See Appendix \ref{subsec:Proof-of-Lem3}. 
\end{IEEEproof}
We are now in position to show the structure of the optimal update
policy. 
\begin{thm}
\label{thm:threshold-structure}For any $s_{1},s_{2}\in\mathcal{S}$,
such that $s_{1}\leq s_{2}$, there is an optimal policy that satisfies
the structural properties as follow:

A) When $\frac{T_{u}}{p_{s}^{T_{u}}}\leq\frac{T_{p}+T_{u}'}{p_{s}^{T_{u}'}}$,
if $\pi^{*}(s_{1})=(0,1)$, then $\pi^{*}(s_{2})=(0,1)$. 

B) When $\frac{T_{u}}{p_{s}^{T_{u}}}\geq\frac{T_{p}+T_{u}'}{p_{s}^{T_{u}'}}$,
if $\pi^{*}(s_{1})=(1,1)$, then $\pi^{*}(s_{2})=(1,1)$. 

\end{thm}
\begin{IEEEproof}
See Appendix \ref{subsec:Proof-of-thm4}. 
\end{IEEEproof}
Theorem \ref{thm:threshold-structure} depicts the structural properties
of the optimal policy $\pi^{*}$ of the SMDP in two cases. We note
that $\frac{T_{u}}{p_{s}^{T_{u}}}$ and $\frac{T_{p}+T_{u}'}{p_{s}^{T_{u}'}}$
can be interpreted as the expected duration of repeatedly taking action
$(0,1)$ and that of taking action $(1,1)$ to get one success transmission,
respectively. Therefore, Theorem 1 also suggests when to choose which
updating action in the high AoI regime. Particularly, in order to
minimize the long-term average cost, the updating action with a shorter
expected duration should be chosen when the AoI is large enough to
dominate the cost. For example, in the first case where the preprocessing
incurs a larger expected update duration, it is better to transmit
the update directly for a large enough AoI, while in the second case
where preprocessing can help shorten the expected update duration,
it is no doubt to choose preprocessing-and-transmission when the AoI
is large enough. The reason why we do not consider the energy consumption
of both actions in the conditions is that the difference between the
energy consumption of different actions is constant and the age increasingly
dominates the cost as the AoI grows larger. We further illustrate
the threshold structure of the optimal policy in Fig. \ref{fig:structure-general},
where the structure of optimal policy falls into case 1 when $v\leq10$,
and otherwise when $v\geq12$.  

\begin{rem}
We note that the result in Theorem \ref{thm:threshold-structure}
can be extended to the case of discrete transmit power control. The
action with the largest transmit power would be the optimal one when
the AoI is large enough, because the largest transmit power can bring
the shortest expected duration. 
\end{rem}
According to Theorem \ref{thm:threshold-structure}, there exists
a threshold $\Omega$ in the optimal update policy. Although the
exact values of $\Omega$ depend on the particular values of $V(s)$,
the structure only depends on the properties of $V(s)$. Therefore,
a low-complexity relative policy iteration algorithm can be developed
by incorporating the threshold structure into a standard relative
policy iteration algorithm. In particular, we will no longer need
to minimize the righthand side of (\ref{eq:bellman-equ}) for all
states to find $\pi^{*}$, thereby reducing the computational complexity.
The details are given in Algorithm \ref{alg:RPI}. 

\begin{algorithm}[t]
\caption{\label{alg:RPI}Relative Policy Iteration based on the Threshold Structure}

\begin{algorithmic}[1]

\STATE \textbf{Initialization:} Set $\pi_{0}^{*}(s)=(0,0)$ for
all $s\in\mathcal{S}$, select a reference state $s^{\dagger}$, and
set $k=0$.

\STATE \textbf{Policy Evaluation:} Given $\pi_{k}^{*}$, compute
the value of $\theta_{k}$ and $V_{k}(s)$ from the linear system
of equations
\begin{equation}
\begin{cases}
\begin{alignedat}{1}\theta_{k}+V_{k}(s)= & \bar{R}(s,\pi_{k}^{*}(s))\\
 & +\sum\limits _{s'\in\mathcal{S}}\bar{p}(s'\mid s,\pi_{k}^{*}(s))V_{k}(s'),
\end{alignedat}
\\
V_{k}(s^{\dagger})=0,
\end{cases}
\end{equation}
 by Gaussian elimination. 

\STATE \textbf{Structured Policy Improvement:} Compute a new policy
$\pi_{k+1}^{*}$ for each $s\in\mathcal{S}$ as follows: 

\textbf{if} $\frac{T_{u}}{p_{s}^{T_{u}}}\leq\frac{T_{p}+T_{u}'}{p_{s}^{T_{u}'}}$
\textbf{and} $\pi_{k+1}^{*}(s-1)=(0,1),$ \textbf{then} $\pi_{k+1}^{*}(s)=(0,1).$ 

\textbf{else if} $\frac{T_{u}}{p_{s}^{T_{u}}}\geq\frac{T_{p}+T_{u}'}{p_{s}^{T_{u}'}}$
\textbf{and} $\pi_{k+1}^{*}(s-1)=(1,1),$ \textbf{then} $\pi_{k+1}^{*}(s)=(1,1).$ 

\textbf{else}
\begin{align}
\pi_{k+1}^{*}(s)=\arg\min_{\boldsymbol{a}\in\mathcal{A}} & \{\bar{R}(s,\pi_{k}^{*}(s))\nonumber \\
 & +\sum\limits _{s'\in\mathcal{S}}\bar{p}(s'\mid s,\pi_{k}^{*}(s))V_{k}(s')\}.
\end{align}

\STATE Set $k=k+1$ and go to Step 2 until $\pi_{k}^{*}(s)=\pi_{k+1}^{*}(s)$
for all $s\in\mathcal{S}$. 

\end{algorithmic}
\end{algorithm}

\section{Special Case Study: Transmission over a Reliable Channel}

In this section, we consider a special scenario where the packets
are transmitted over a reliable channel. Accordingly, the status updating
problem can be simplified. From (\ref{eq:Dynamic}) we can see that
the states smaller than $\min\{T_{u},T_{p}+T_{u}'\}$ are non-recurrent
states. Since the policies in non-recurrent states has no effect
on the average cost, we can only consider the state space $\mathcal{S}^{\dagger}\triangleq\left\{ \min\{T_{u},T_{p}+T_{u}'\},\cdots,\hat{\delta}\right\} $
when discussing the optimal policy. 

\subsection{Case 1}

\begin{figure}[tp]
\centering

\includegraphics[scale=0.55]{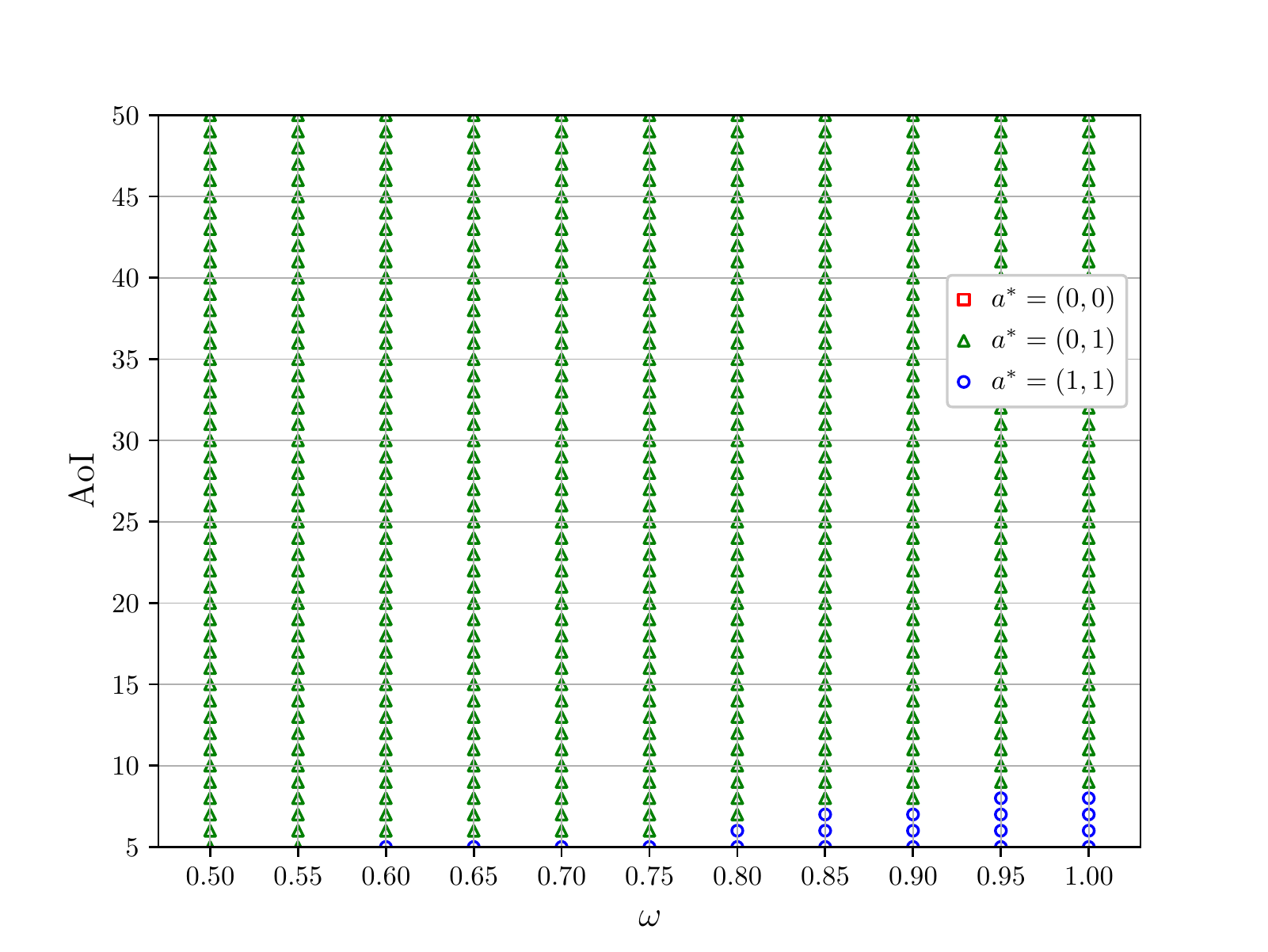}

\caption{\label{fig:structure-sp1}Structure of the optimal policy in Theorem
\ref{thm:threshold1} for different values of $\omega$ ($T_{u}=5$,
$T_{u}'=1$, $l=3$, $v=5$, $f=15$, $\tau=1$, $\kappa=0.00005$,
$P=3$).}
\end{figure}

Based on the model of the reliable channel, we give the first simplification
of the optimal policy. 
\begin{lem}
\label{lem:simplify1}For any $s\in\mathcal{S}^{\dagger}$, we have
$\pi^{*}(s)\neq(0,0)$ when $\frac{1}{2}L(\boldsymbol{a}_{f})(L(\boldsymbol{a}_{f})+1)\geq\omega C(\boldsymbol{a}_{f})$. 
\end{lem}
\begin{IEEEproof}
See Appendix \ref{subsec:Proof-of-lem:simplify1}. 
\end{IEEEproof}
Lemma \ref{lem:simplify1} indicates that the IoT device will never
stay idle with the optimal policy when the AoI dominates in the cost.
Accordingly, the threshold structure in Theorem \ref{thm:threshold-structure}
can be simplified, which is presented in the theorem below.
\begin{thm}
\label{thm:threshold1}For $s\in\mathcal{S}^{\dagger}$, the optimal
policy is of a switch-type structure when $\frac{1}{2}L(\boldsymbol{a}_{f})(L(\boldsymbol{a}_{f})+1)\geq\omega C(\boldsymbol{a}_{f})$,
namely, there exists a threshold $\Omega\geq\min\{T_{u},T_{p}+T_{u}'\}$,
such that when $T_{u}\leq T_{p}+T_{u}'$,
\begin{equation}
\pi^{*}(s)=\begin{cases}
(1,1), & T_{u}\leq s<\Omega,\\
(0,1), & s\geq\Omega,
\end{cases}\label{eq:switching-structure-1-2}
\end{equation}
and when $T_{u}\geq T_{p}+T_{u}'$, 
\begin{equation}
\pi^{*}(s)=\begin{cases}
(0,1), & T_{p}+T_{u}'\leq s<\Omega,\\
(1,1), & s\geq\Omega.
\end{cases}\label{eq:switching-structure-1-1}
\end{equation}
\end{thm}
\begin{IEEEproof}
According to Lemma \ref{lem:simplify1}, we can exclude action $(0,0)$
from the optimal policy when $\frac{1}{2}L(\boldsymbol{a}_{f})(L(\boldsymbol{a}_{f})+1)\geq\omega C(\boldsymbol{a}_{f})$.
Moreover, since we have proved the threshold structure of the optimal
policy in a general case in Theorem \ref{thm:threshold-structure},
the optimal policy can be further proved to satisfy the switching
structure in (\ref{eq:switching-structure-1-2}) and (\ref{eq:switching-structure-1-1}).
\end{IEEEproof}
Theorem \ref{thm:threshold1} depicts the structure  of the optimal
policy $\pi^{*}$ for the SMDP in (\ref{eq:Problem}) when $p_{s}=1$
and $\frac{1}{2}L(\boldsymbol{a}_{f})(L(\boldsymbol{a}_{f})+1)\geq\omega C(\boldsymbol{a}_{f})$.
We further illustrate the analytical results of Theorem \ref{thm:threshold1}
in Fig. \ref{fig:structure-sp1}, where $T_{u}\leq T_{p}+T_{u}'$
and $C(\bm{a}=(0,1))>C(\bm{a}=(1,1))$. It can be seen from Fig. \ref{fig:structure-sp1}
that the optimal policy is of the threshold type. Moreover, the threshold
increases along with $\omega$. This indicates that, when the weighting
factor is large, it is not desirable to directly transmit a new status
update to the destination due to a high weighted energy consumption.

\begin{figure}[tp]
\centering

\subfloat[]{\includegraphics[width=0.5\textwidth]{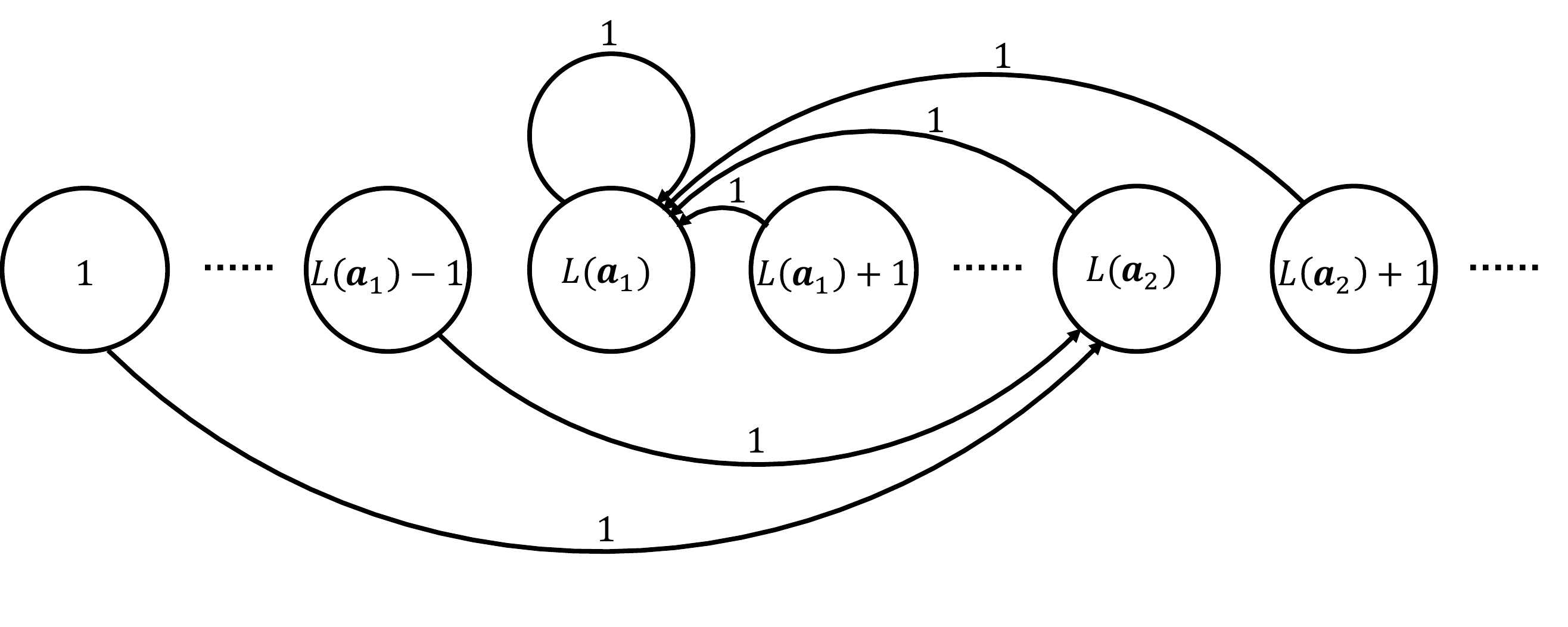}}

\subfloat[]{\includegraphics[width=0.5\textwidth]{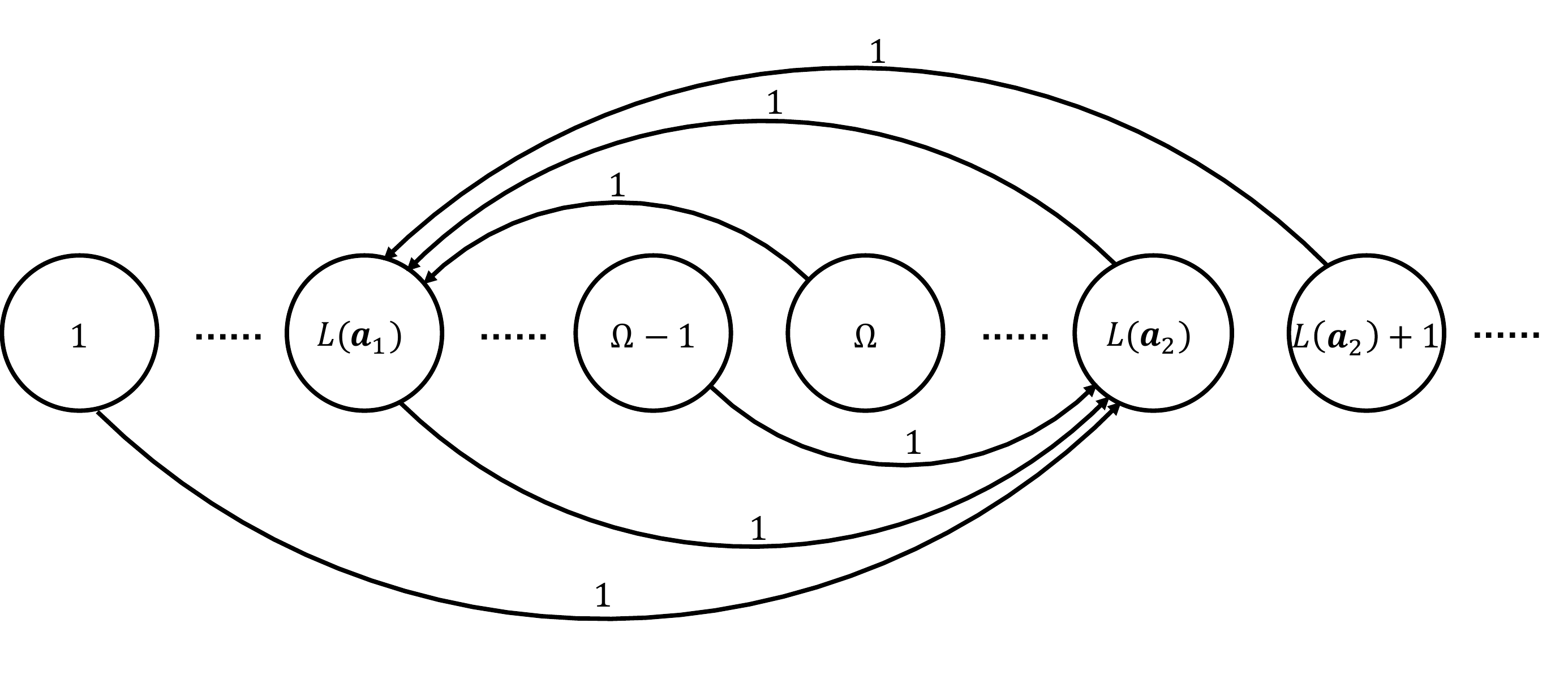}}

\subfloat[]{\includegraphics[width=0.5\textwidth]{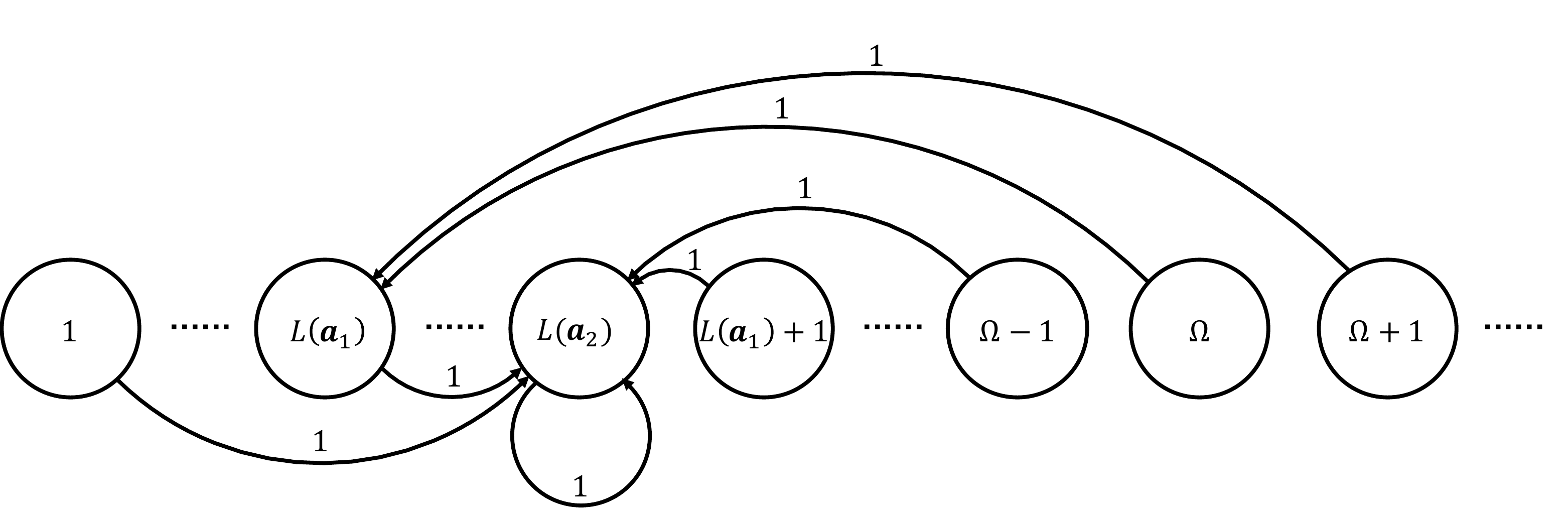}}\caption{\label{fig:MarkovChain-1}The states transitions under a threshold
policy with different values of $\Omega$. (a) $\Omega=L(\boldsymbol{a}_{1})$.
(b) $L(\boldsymbol{a}_{1})<\Omega\protect\leq L(\boldsymbol{a}_{2})$.
(c) $\Omega>L(\boldsymbol{a}_{2})$.}
\end{figure}

We then denote $\boldsymbol{a}_{1}=\arg\min\limits _{\boldsymbol{a}\in\mathcal{A}\setminus(0,0)}\{L(\boldsymbol{a})\}$
and $\boldsymbol{a}_{2}=\arg\max\limits _{\boldsymbol{a}\in\mathcal{A}\setminus(0,0)}\{L(\boldsymbol{a})\}$.
According to the threshold structure in Theorem \ref{thm:threshold1},
we can proceed to reduce the recurrent state space of the computing-enable
IoT system. 
\begin{lem}
\label{lem:recurrent-states}For a given threshold policy of the type
in Theorem \ref{thm:threshold1} with the threshold of $\Omega$,
recurrent state space $\mathcal{S}'$ can be given as follow: 

A) $\mathcal{S}'=\left\{ L(\boldsymbol{a}_{1})\right\} $ when $\Omega=L(\boldsymbol{a}_{1})$. 

B) $\mathcal{S}'=\left\{ L(\boldsymbol{a}_{1}),L(\boldsymbol{a}_{2})\right\} $
when $L(\boldsymbol{a}_{1})<\Omega\leq L(\boldsymbol{a}_{2})$. 

C) $\mathcal{S}'=\left\{ L(\boldsymbol{a}_{2})\right\} $ when $\Omega>L(\boldsymbol{a}_{2})$. 
\end{lem}
\begin{IEEEproof}
As illustrated in Fig. \ref{fig:MarkovChain-1}, we can use a Discrete
Time Markov Chain (DTMC) to model the MDP constructed by any threshold
policy of the type in Theorem \ref{thm:threshold1} with the threshold
of $\Omega$. It can be seen from Fig. \ref{fig:MarkovChain-1}(a)
and Fig. \ref{fig:MarkovChain-1}(c), respectively, that there is
only one recurrent state $L(\boldsymbol{a}_{1})$ when $\Omega=L(\boldsymbol{a}_{1})$
and $L(\boldsymbol{a}_{2})$ when $\Omega>L(\boldsymbol{a}_{2})$.
Also we can see from Fig. \ref{fig:MarkovChain-1}(b) that the recurrent
states are $L(\boldsymbol{a}_{1})$ and $L(\boldsymbol{a}_{2})$ when
$L(\boldsymbol{a}_{1})<\Omega\leq L(\boldsymbol{a}_{2})$. 
\end{IEEEproof}

Under the threshold policy, we proceed with analyzing the average
cost of any threshold $\Omega$. 
\begin{lem}
\label{lem:threshold-range}Let $J_{1}=\frac{3}{2}L(\boldsymbol{a}_{1})+\omega\frac{C(\boldsymbol{a}_{1})}{L(\boldsymbol{a}_{1})}-\frac{1}{2}$,
$J_{2}=L(\boldsymbol{a}_{1})L(\boldsymbol{a}_{2})+\omega\frac{C(\boldsymbol{a}_{1})+C(\boldsymbol{a}_{2})}{L(\boldsymbol{a}_{1})+L(\boldsymbol{a}_{2})}$,
$J_{3}=\frac{3}{2}L(\boldsymbol{a}_{2})+\omega\frac{C(\boldsymbol{a}_{2})}{L(\boldsymbol{a}_{2})}-\frac{1}{2}$.
For a given threshold $\Omega$, the average cost of the threshold
policy in Theorem \ref{thm:threshold1} is given by
\begin{equation}
J(\Omega)=\begin{cases}
J_{1}, & \Omega=L(\boldsymbol{a}_{1}),\\
J_{2}, & L(\boldsymbol{a}_{1})<\Omega\leq L(\boldsymbol{a}_{2}),\\
J_{3}, & \Omega>L(\boldsymbol{a}_{2}).
\end{cases}
\end{equation}
\end{lem}
\begin{IEEEproof}
See Appendix \ref{subsec:Proof-of-threshold-range}. 
\end{IEEEproof}
By leveraging the above results, we can find the set of the optimal
threshold $\Omega^{*}$. 
\begin{thm}
\label{thm:opt-threshold-range}If $J_{1}$ is smaller than $J_{2}$
and $J_{3}$, we have $\Omega^{*}=L(\boldsymbol{a}_{1})$. If $J_{2}$
is smaller than $J_{1}$ and $J_{3}$, we have $L(\boldsymbol{a}_{1})<\Omega^{*}\leq L(\boldsymbol{a}_{2})$.
If $J_{3}$ is smaller than $J_{1}$ and $J_{2}$, we have $\Omega^{*}>L(\boldsymbol{a}_{2})$. 
\end{thm}
\begin{IEEEproof}
According to Lemma \ref{lem:threshold-range}, we can determine the
set of the optimal threshold $\Omega^{*}$ by comparing the values
of $J_{1}$, $J_{2}$ and $J_{3}$. 
\end{IEEEproof}
We have proved in Lemma \ref{lem:recurrent-states} that under the
threshold policy in Theorem \ref{thm:threshold1}, only a few states
of the system are recurrent states. Once the set of the optimal threshold
is determined, we can determine the optimal policy in the recurrent
states. Therefore, the specific value of the threshold is not a necessity.
As long as we take any value in the set of the optimal threshold,
we can achieve the goal of minimizing the average cost. 

\subsection{Case 2}

\begin{figure}[tp]
\centering

\includegraphics[scale=0.55]{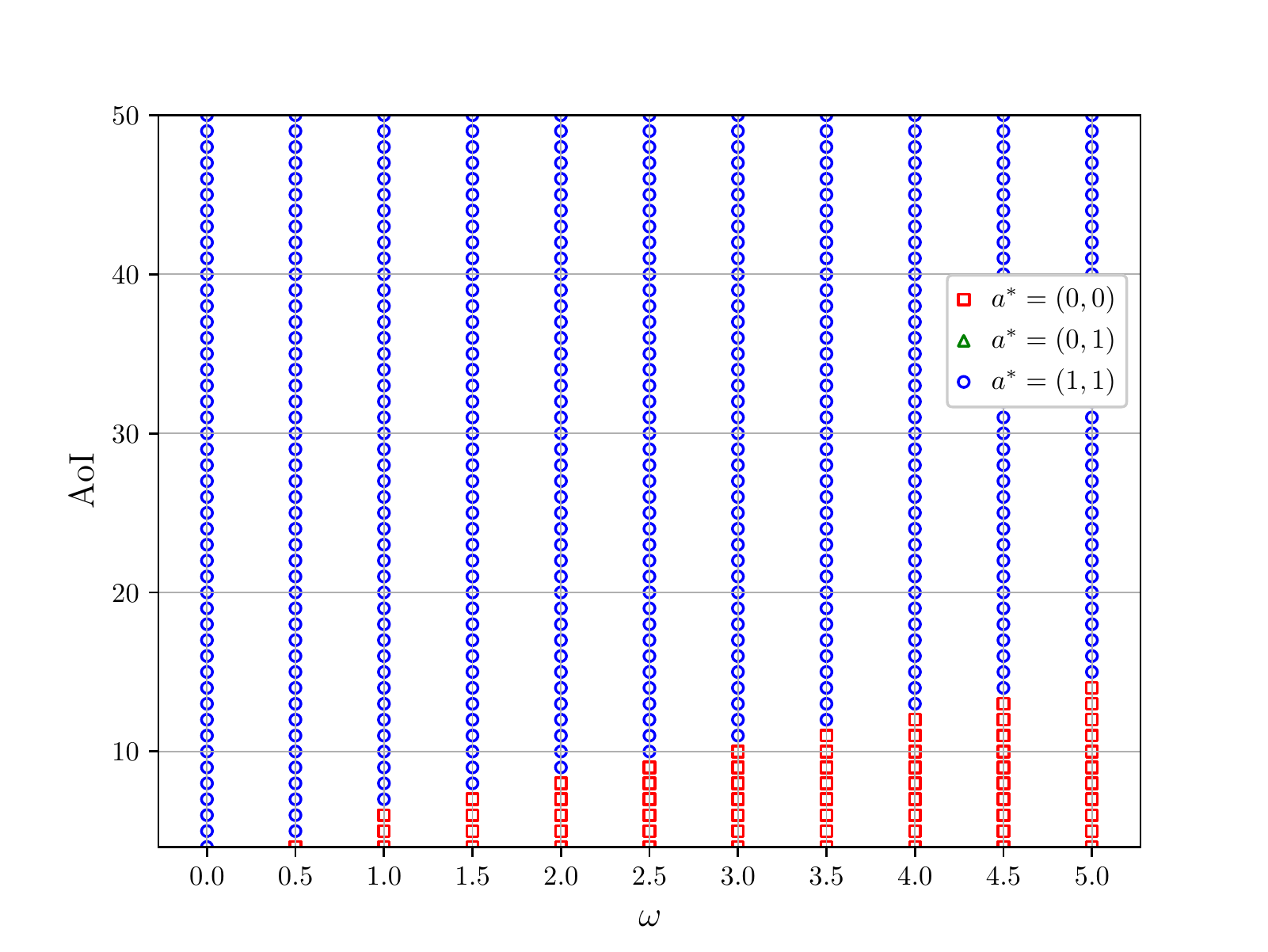}

\caption{\label{fig:structure-sp2}Structure of the optimal policy in Theorem
\ref{thm:threshold2} for different values of $\omega$ ($T_{u}=6$,
$T_{u}'=2$, $l=3$, $v=5$, $f=45$, $\tau=1$, $\kappa=0.00005$,
$P=6$).}
\end{figure}

Based on the model of the reliable channel, we give the second simplification
of the optimal policy. Recall that $\boldsymbol{a}_{1}=\arg\min\limits _{\boldsymbol{a}\in\mathcal{A}\setminus(0,0)}\{L(\boldsymbol{a})\}$
and $\boldsymbol{a}_{2}=\arg\max\limits _{\boldsymbol{a}\in\mathcal{A}\setminus(0,0)}\{L(\boldsymbol{a})\}$,
we have the following lemma.
\begin{lem}
\label{lem:simplify2}For any $s\in\mathcal{S}^{\dagger}$, we have
$\pi^{*}(s)\neq\boldsymbol{a}_{2}$ when $\frac{C(\boldsymbol{a}_{1})}{L(\boldsymbol{a}_{1})}\leq\frac{C(\boldsymbol{a}_{2})}{L(\boldsymbol{a}_{2})}$. 
\end{lem}
\begin{IEEEproof}
See Appendix \ref{subsec:Proof-of-simplify2}. 
\end{IEEEproof}
Lemma \ref{lem:simplify2} reveals that the action with a lower energy
efficiency (i.e., a larger energy consumption per minislot) shall
be excluded in the optimal policy. Accordingly, the threshold structure
in Theorem \ref{thm:threshold-structure} is simplified in the following
theorem. 
\begin{thm}
\label{thm:threshold2}For $s\in\mathcal{S}^{\dagger}$, the optimal
policy is of a switch-type structure when $\frac{C(\boldsymbol{a}_{1})}{L(\boldsymbol{a}_{1})}\leq\frac{C(\boldsymbol{a}_{2})}{L(\boldsymbol{a}_{2})}$,
namely, there is a threshold $\Omega\geq\min\{T_{u},T_{p}+T_{u}'\}$,
such that
\begin{equation}
\pi^{*}(s)=\begin{cases}
(0,0), & L(\boldsymbol{a}_{1})\leq s<\Omega,\\
\boldsymbol{a}_{1}, & s\geq\Omega.
\end{cases}\label{eq:switching-structure-2}
\end{equation}
\end{thm}
\begin{IEEEproof}
According to Lemma \ref{lem:simplify2}, we can exclude $\bm{a}_{2}$
from the optimal policy when $\frac{C(\boldsymbol{a}_{1})}{L(\boldsymbol{a}_{1})}\leq\frac{C(\boldsymbol{a}_{2})}{L(\boldsymbol{a}_{2})}$.
Moreover, since we have proved the threshold structure of the optimal
policy in a general case in Theorem \ref{thm:threshold-structure},
the optimal policy can be further proved to satisfy the switching
structure in (\ref{eq:switching-structure-2}). 
\end{IEEEproof}
Theorem \ref{thm:threshold2} depicts the structure of the optimal
policy $\pi^{*}$ for the SMDP in (\ref{eq:Problem}) when $p_{s}=1$
and $\frac{C(\boldsymbol{a}_{1})}{L(\boldsymbol{a}_{1})}\leq\frac{C(\boldsymbol{a}_{2})}{L(\boldsymbol{a}_{2})}$.
Fig. \ref{fig:structure-sp2} illustrates the analytical results of
Theorem \ref{thm:threshold2}, where $p_{s}=1$ and $C_{u}\geq\frac{T_{p}C_{p}+T_{u}'C_{u}}{T_{p}+T_{u}'}$.
We can see from Fig. \ref{fig:structure-sp2} that, in order to strike
a balance between the AoI and the energy consumption, the IoT device
does not transmit until the AoI is large. We can also see that the
threshold increases with the increasing of the weighting factor $\omega$.
This is due to a higher weighted energy consumption in average cost
when $\omega$ grows larger. 

Under the threshold policy in Theorem \ref{thm:threshold2}, we proceed
with analyzing the average cost. 
\begin{lem}
\label{lem:threshold-value}The average cost of the threshold policy
for any given threshold $\Omega$ in Theorem \ref{thm:threshold2}
can be given by 
\begin{equation}
J(\Omega)=L(\boldsymbol{a}_{1})+\frac{1}{2}(\Omega-1)+\frac{\omega C(\bm{a}_{1})}{\Omega}.
\end{equation}
\end{lem}
\begin{IEEEproof}
See Appendix \ref{subsec:Proof-of-threshold-value}. 
\end{IEEEproof}
By leveraging the above results, we can find the optimal threshold
value $\Omega^{*}$. 
\begin{thm}
\label{thm:opt-threshold2}The optimal threshold $\Omega^{*}$ of
the optimal update policy in Theorem \ref{thm:threshold2} is given
by 
\begin{equation}
\Omega^{*}=\arg\min\left(J\left(\left\lfloor \sqrt{2\omega C(\bm{a}_{1})}\right\rfloor \right),J\left(\left\lceil \sqrt{2\omega C(\bm{a}_{1})}\right\rceil \right)\right).\label{eq:opt-threshold}
\end{equation}
\end{thm}
\begin{IEEEproof}
See Appendix \ref{subsec:Proof-of-opt-threshold}. 
\end{IEEEproof}
From the expression of the optimal threshold $\Omega^{*}$, we can
see that the threshold is monotonically increasing with respect to
$\omega$ and $C(\boldsymbol{a}_{1})$, which is consistent with the
result in Fig. \ref{fig:structure-sp2}. This indicates that, when
the weighting factor or the energy consumption is large, transmitting
a new status update can achieve better result than keeping idle only
in the large AoI regime. 
\begin{rem}
In this section, by studying the special cases, we show that the optimal
policy has a switching structure with respect to only two actions.
Moreover, the optimal threshold can be obtained in closed-form, which
reveals how the system parameters affect the threshold policy. It
is worth noting that, in practice, the conclusions obtained from the
special cases study can be applied in the high SNR regime and the
obtained optimal policy can be used as an approximation of the optimal
policy when the success rate is high.
\end{rem}

\section{Simulation Results}

In this section, we present the simulation results to explore the
effects of system parameters on the optimal update policy and demonstrate
the efficacy of the proposed scheme by comparing it with two other
zero-wait policies, i.e., zero-wait no-computation policy and zero-wait
computation policy. Particularly, in both baseline policies, the IoT
device starts a new transmission immediately after the previous transmission
is finished. In the zero-wait no-computation policy, the IoT device
transmits each status update without preprocessing, while, in the
zero-wait computation policy, the IoT device preprocesses each status
update and then transmits the processed status update. In the simulations,
we truncate the state space by setting the upper limit of the number
of states to be 200.

\subsection{Performance Evaluation in the General Case}

\begin{figure}[tp]
\centering

\subfloat[\label{fig:ps1}]{\includegraphics[width=0.5\textwidth]{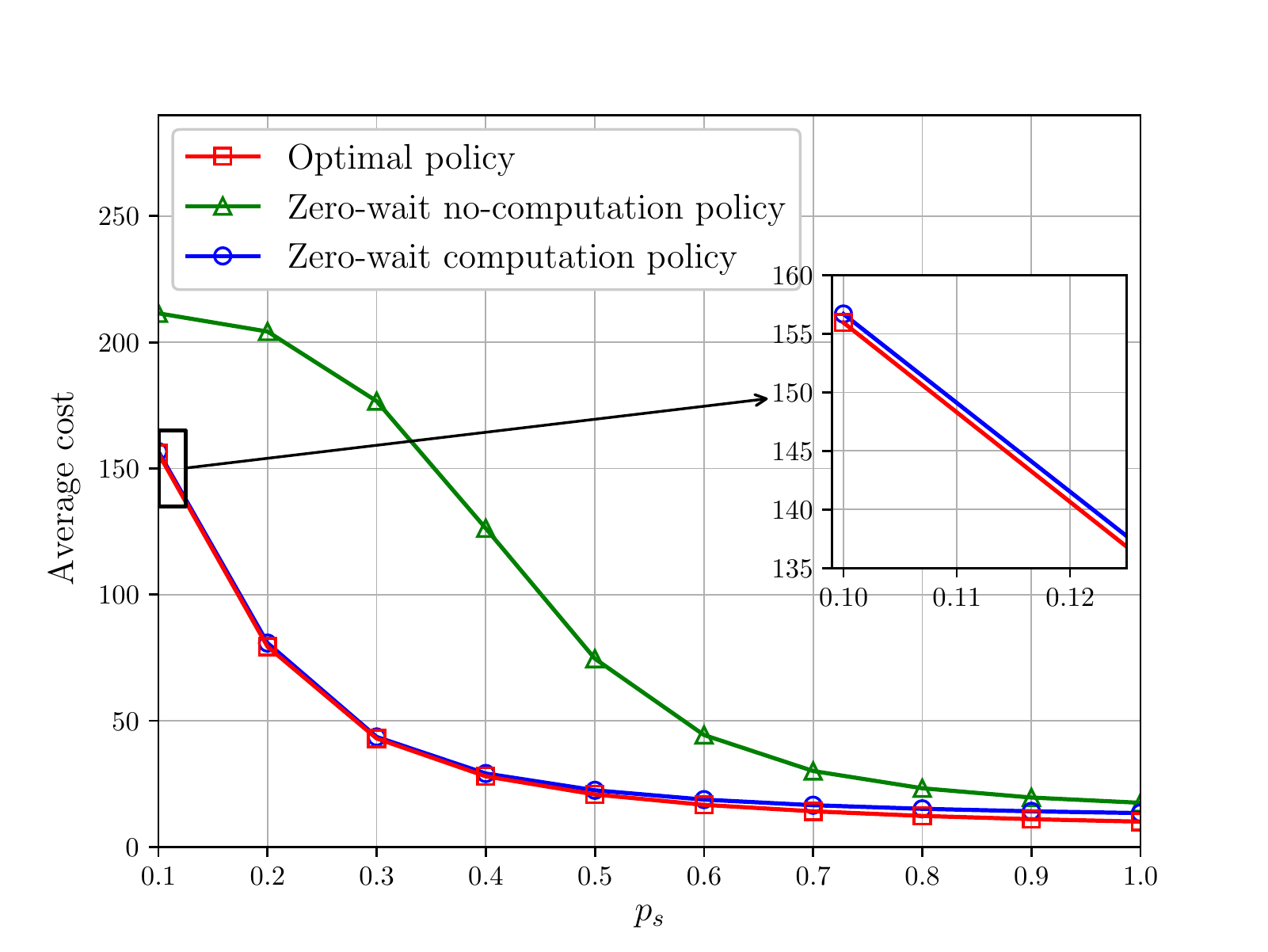}}

\subfloat[\label{fig:ps2}]{\includegraphics[width=0.5\textwidth]{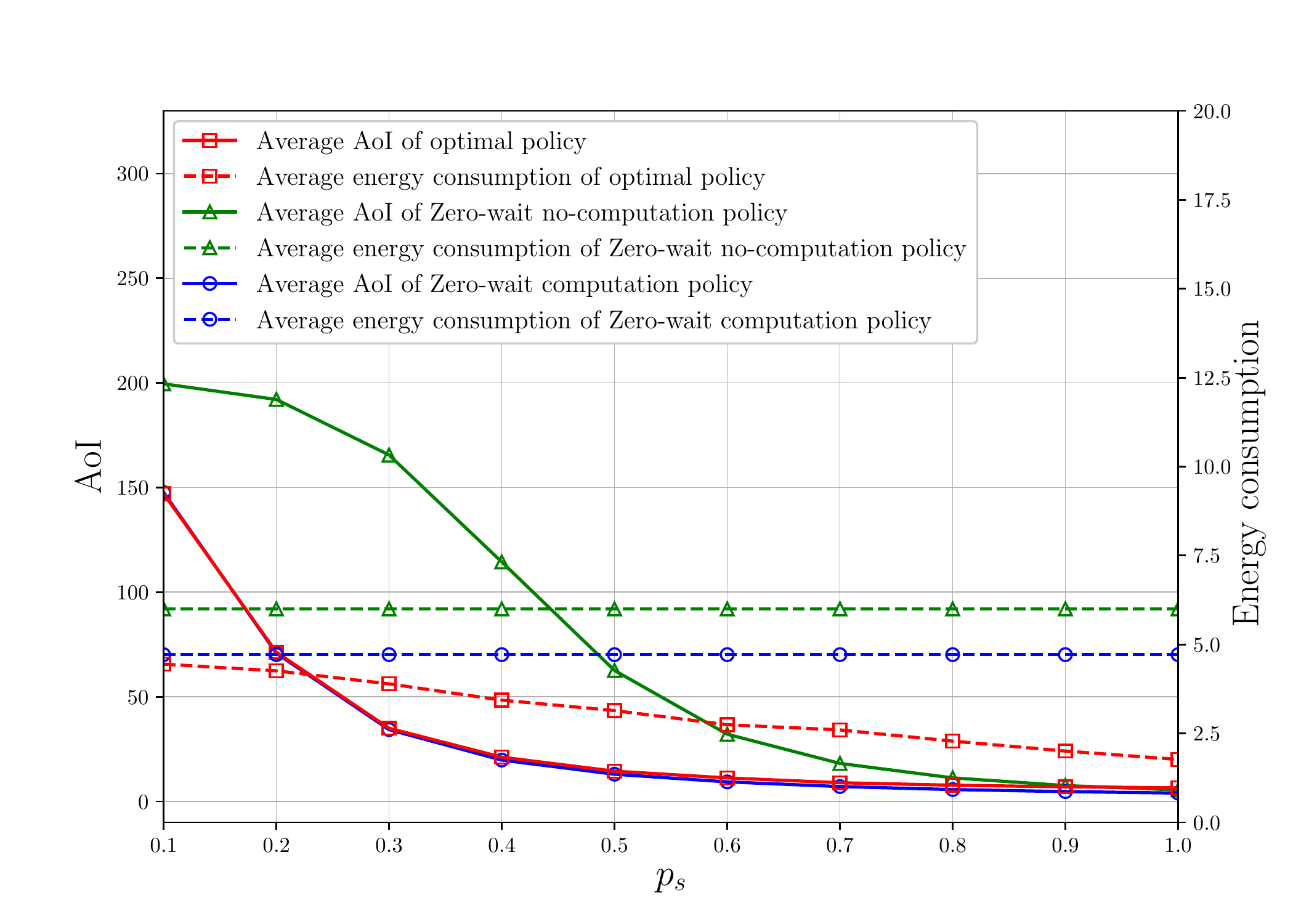}}\caption{\label{fig:opt-vers-zerowait-ps}Performance comparison among the
optimal policy, the zero-wait no-computation policy, and the zero-wait
computation policy ($T_{u}=4$, $T_{u}'=2$, $l=3$, $v=2$, $f=35$,
$\tau=1$, $\kappa=0.00005$, $P=6$, $\omega=2$). (a) The average
cost versus $p_{s}$. (b) The average AoI and the average energy consumption
versus $p_{s}$. }
\end{figure}

In Fig. \ref{fig:opt-vers-zerowait-ps}, the average cost of the optimal
policy and the two baseline policies are compared with respect to
the transmission success probability $p_{s}$. As we can see from
Fig. \ref{fig:opt-vers-zerowait-ps}(a), the optimal policy outperforms
the zero-wait policies. Moreover, the average cost decreases as $p_{s}$
increases. The reason can be explained with the aid of Fig \ref{fig:opt-vers-zerowait-ps}(b).
First, it is evident that the average AoI of all the three policies
decreases with the increasing of $p_{s}$ since the AoI is more likely
to be reset with a larger transmission success probability. Second,
because the IoT device with zero-wait policies keep updating continuously,
the average energy consumption remains a constant irrespective of
$p_{s}$.  For $0.1\leq p_{s}\leq1$, we can see that the average
energy consumption steadily decreases with the increase of $p_{s}$.
This is due to the fact that less transmission is needed to reduce
the AoI when $p_{s}$ is larger. Therefore, the optimal policy can
adapt to the channel quality. Through this comparison, we can see
that although the optimal update policy does not yield the minimum
AoI, it has a smaller energy consumption than the zero-wait policies.
By trading off the AoI for energy consumption, the optimal update
policy achieves the smallest average cost. 

\begin{figure}[tp]
\centering

\subfloat[\label{fig:v1}]{\includegraphics[width=0.5\textwidth]{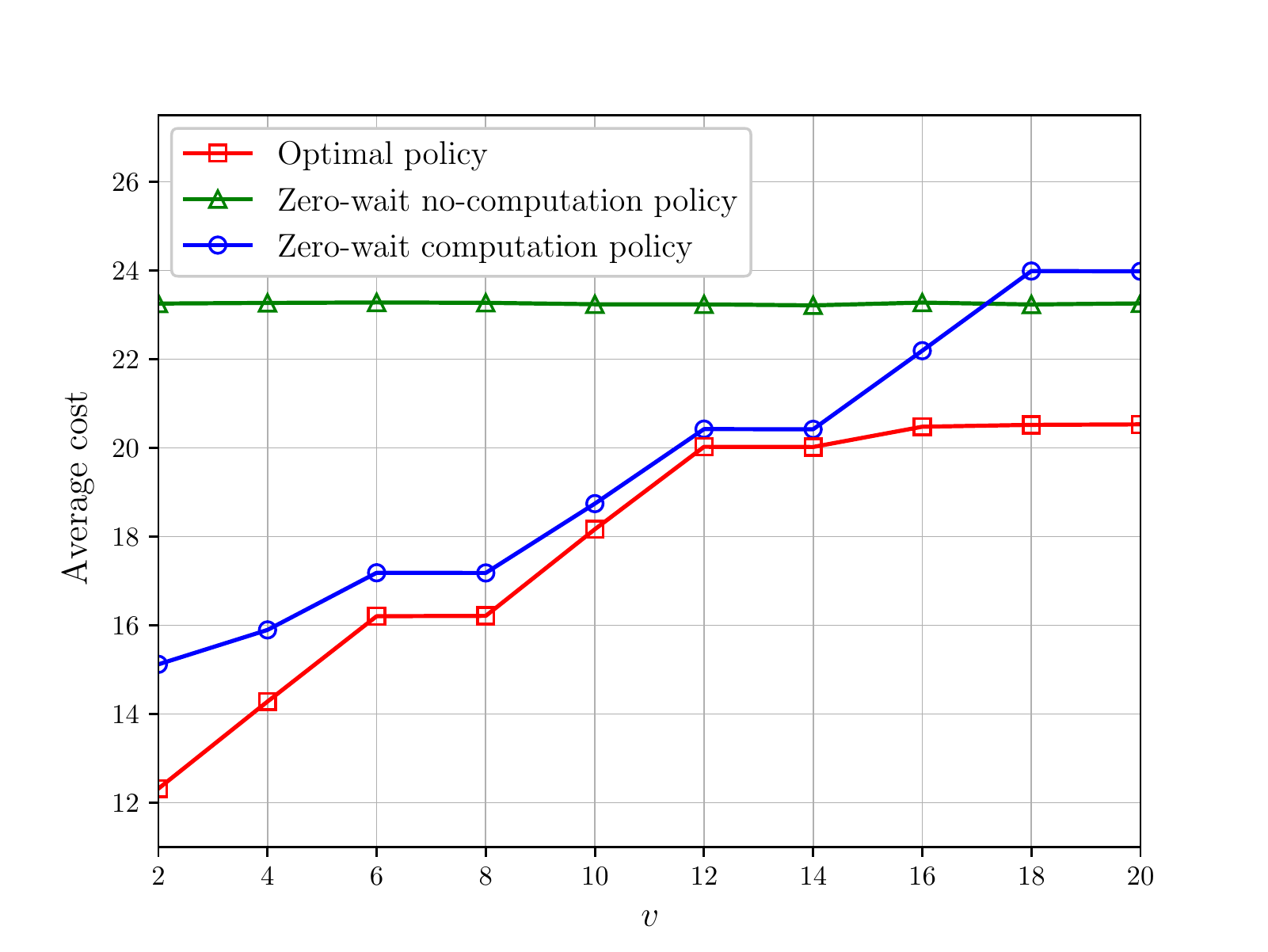}}

\subfloat[\label{fig:v2}]{\includegraphics[width=0.5\textwidth]{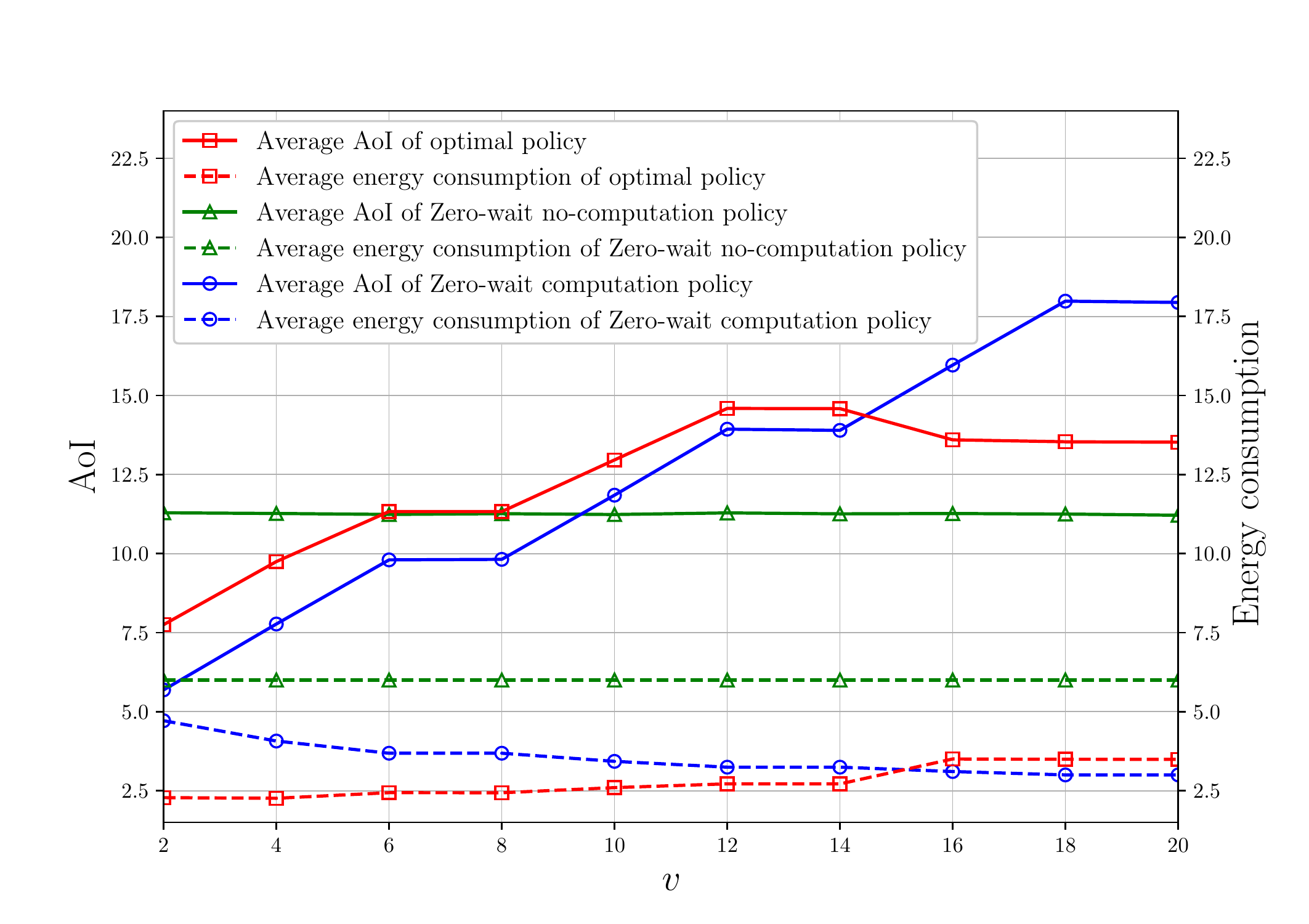}}\caption{\label{fig:opt-vers-zerowait-v}Performance comparison among the optimal
policy, the zero-wait no-computation policy, and the zero-wait computation
policy ($T_{u}=4$, $T_{u}'=2$, $l=3$, $f=35$, $\tau=1$, $\kappa=0.00005$,
$P=6$, $\omega=2$, $p_{s}=0.8$). (a) The average cost versus $v$.
(b) The average AoI and the average energy consumption versus $v$.}
\end{figure}

In Fig. \ref{fig:opt-vers-zerowait-v}, the average cost of the optimal
policy and the two baseline policies are compared with respect to
the number of CPU cycles required to preprocess one bit $v$. We can
see from Fig. \ref{fig:opt-vers-zerowait-v}(a) that the optimal policy
outperforms both baseline policies. It is easy to see that the performance
of zero-wait no-computation policy is irrespective of $v$ since each
status update is transmitted directly without preprocessing. In contrast,
the average cost of the optimal policy and the zero-wait computation
policy is non-decreasing as $v$ grows. Particularly, we can see from
Fig. \ref{fig:opt-vers-zerowait-v}(b) that the average AoI of the
zero-wait computation policy increases with $v$. Since the status
update is more and more computation intensive as $v$ grows, the preprocessing
at the IoT device requires more and more time, which leads to an increase
in the average AoI. However, since the computation energy consumption
per minislot in this setup is less than the transmission energy consumption
per minislot, the average energy consumption of the zero-wait computation
policy is shown to decline with $v$. In other words, although the
duration and total energy consumption are increasing as $v$ increases,
the average energy consumption is decreasing. Moreover, we can see
from Fig. \ref{fig:opt-vers-zerowait-v}(b) that the average AoI of
the optimal update policy is non-decreasing with $v$ except when
$14\leq v\leq16$. The reason why the average AoI of the optimal policy
drops for $14\leq v\leq16$ can be explained with Fig. \ref{fig:structure-general}.
Since $v=16$, the optimal policy completely abandons the action of
$(1,1)$. The change of the optimal policy effectively reduces the
average AoI of the system, but induces a sudden increase in the average
energy consumption at $v=16$. We can also see that the average energy
consumption increases as $v$ increases. This is because the optimal
policy makes sacrifices in energy consumption in order to ensure that
the AoI is increased at a slower pace. It could be concluded that
the optimal update policy can adjust adaptively based on the degree
of computational intensity of the status update. 

\subsection{Performance Evaluation in the Special Cases}

\begin{figure}[tp]
\centering

\subfloat[\label{fig:sp1-weight1}]{\includegraphics[width=0.5\textwidth]{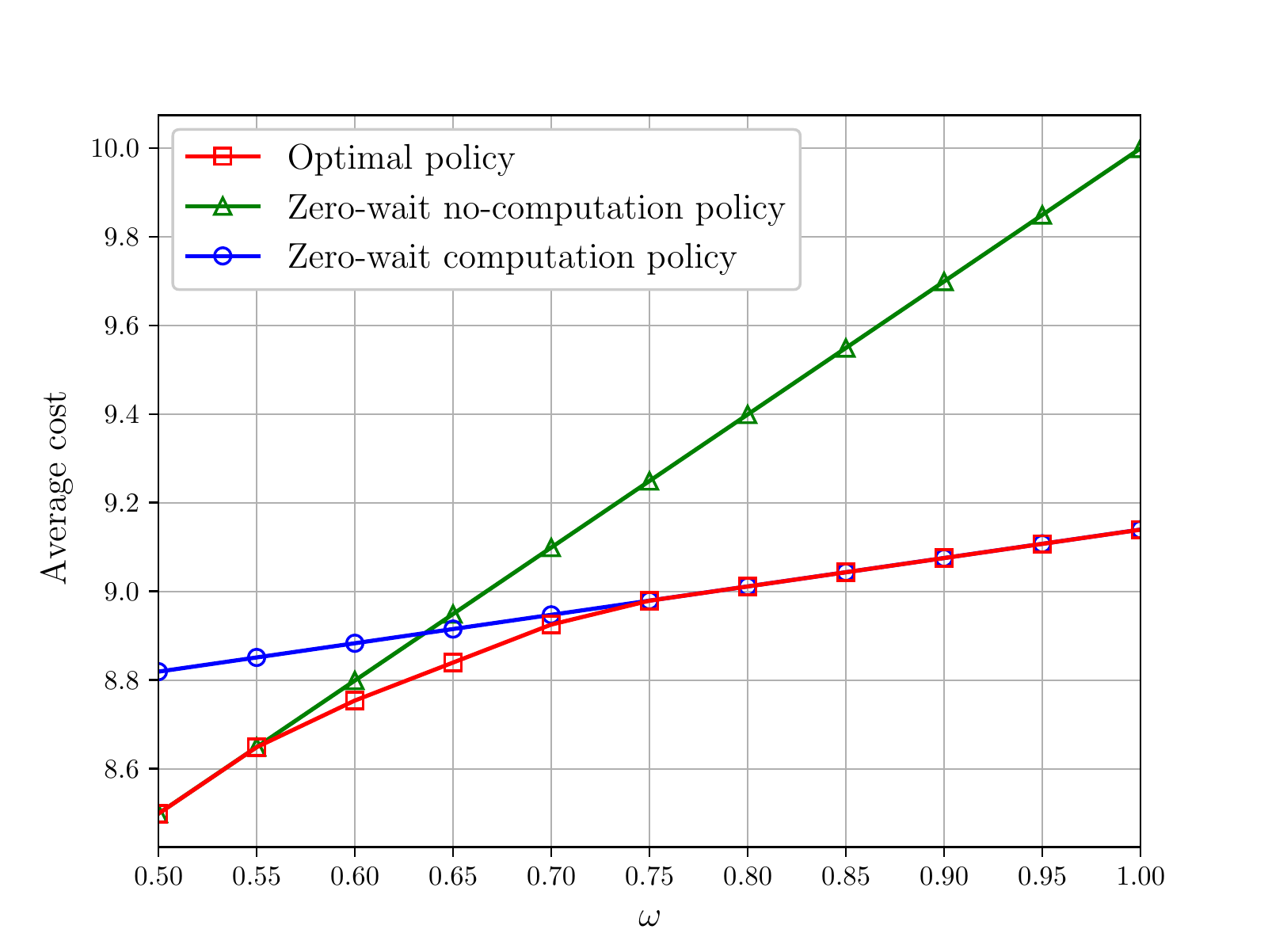}}

\subfloat[\label{fig:sp1-weight2}]{\includegraphics[width=0.5\textwidth]{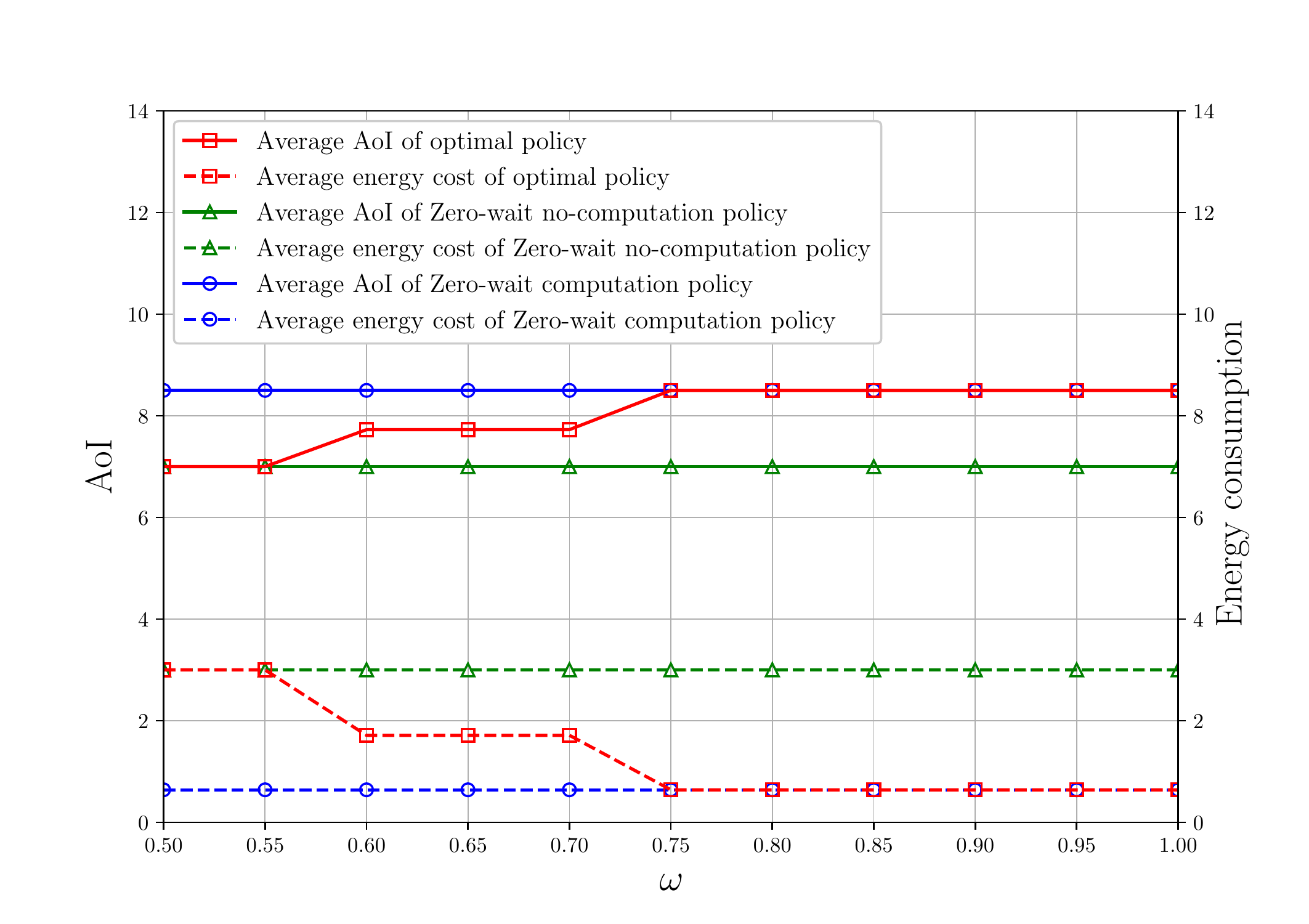}}\caption{\label{fig:opt-vers-zerowait-sp1-weight}Performance comparison among
the optimal policy, the zero-wait no-computation policy, and the zero-wait
computation policy ($T_{u}=5$, $T_{u}'=1$, $l=3$, $v=5$, $f=15$,
$\tau=1$, $\kappa=0.00005$, $P=3$). (a) The average cost versus
$\text{\ensuremath{\omega}}$. (b) The average AoI and the average
energy consumption versus $\omega$.}
\end{figure}

In this subsection, we show the performance of two different special
cases in Section IV that the packets are transmitted over a reliable
channel. In Fig. \ref{fig:opt-vers-zerowait-sp1-weight}, the average
cost of the optimal policy and the two baseline policies are compared
with respect to the weighting factor $\omega$ when $T_{u}\leq T_{p}+T_{u}'$
and $\frac{1}{2}(T_{p}+T_{u}')(T_{p}+T_{u}'+1)\geq\omega(T_{p}C_{p}+T_{u}'C_{u})$.
As we can see from Fig. \ref{fig:opt-vers-zerowait-sp1-weight}(a),
the optimal policy outperforms both two baseline policies.  Moreover,
the optimal policy coincides with the zero-wait no-computation policy
when $\omega$ is small and coincides with the zero-wait computation
policy when $\omega$ is large. This is because, in this simulation
setup, when $w\leq0.55$, we have $\min(J_{1},J_{2},J_{3})=J_{1}$
and the optimal policy is to always transmit the status update directly,
while when $w\ge0.75$, we have $\min(J_{1},J_{2},J_{3})=J_{3}$ and
the optimal policy is to always preprocess and transmit the status
update. Moreover, when $0.60\leq w\leq0.70$, we have $\min(J_{1},J_{2},J_{3})=J_{2}$
and the optimal policy is to execute the above two actions in turns.
That is the reason why the optimal policy outperforms the two zero-wait
policies in this regime. 

\begin{figure}[tp]
\centering

\subfloat[\label{fig:sp2-weight1}]{\includegraphics[width=0.5\textwidth]{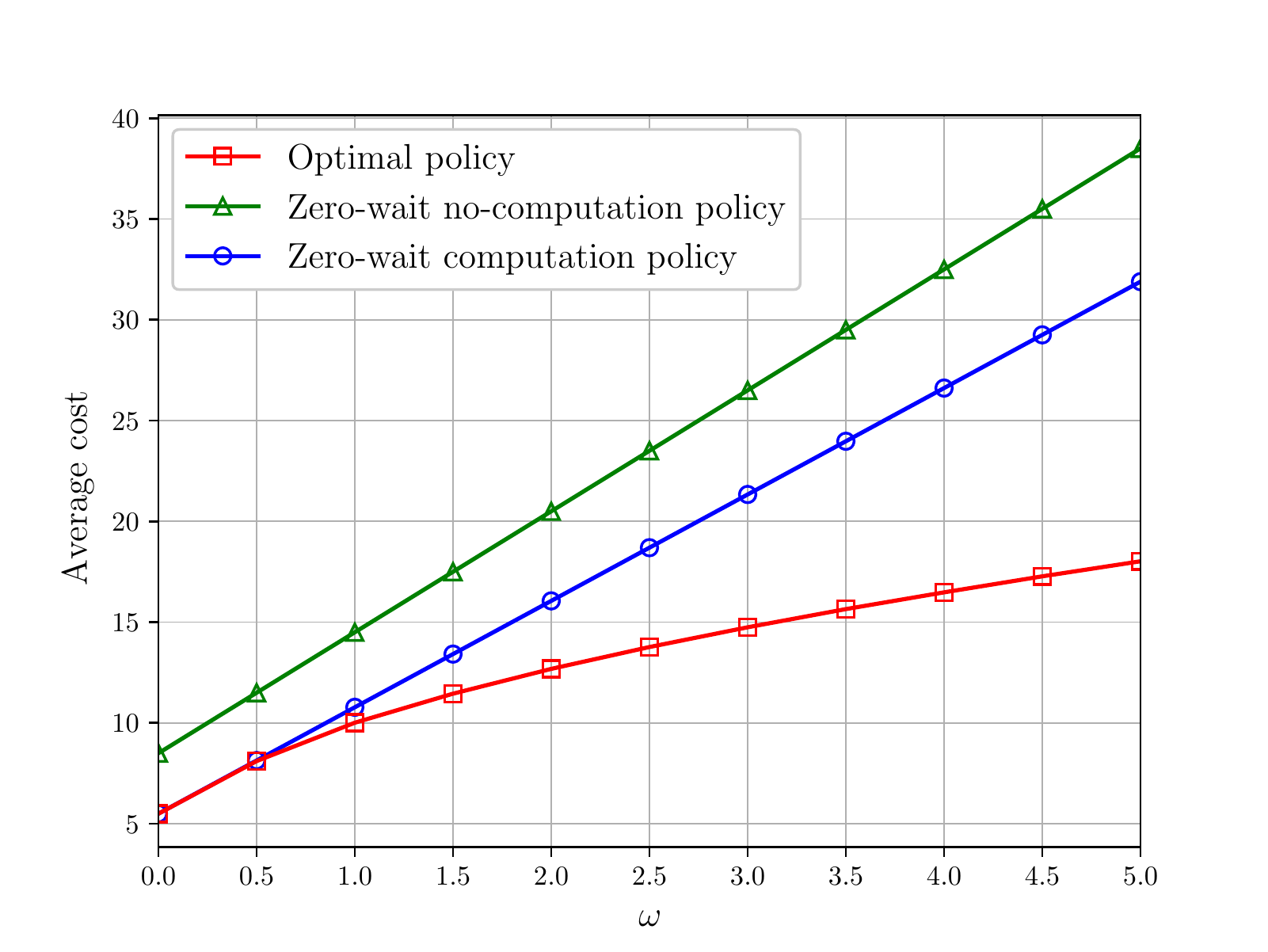}}

\subfloat[\label{fig:sp2-weight2}]{\includegraphics[width=0.5\textwidth]{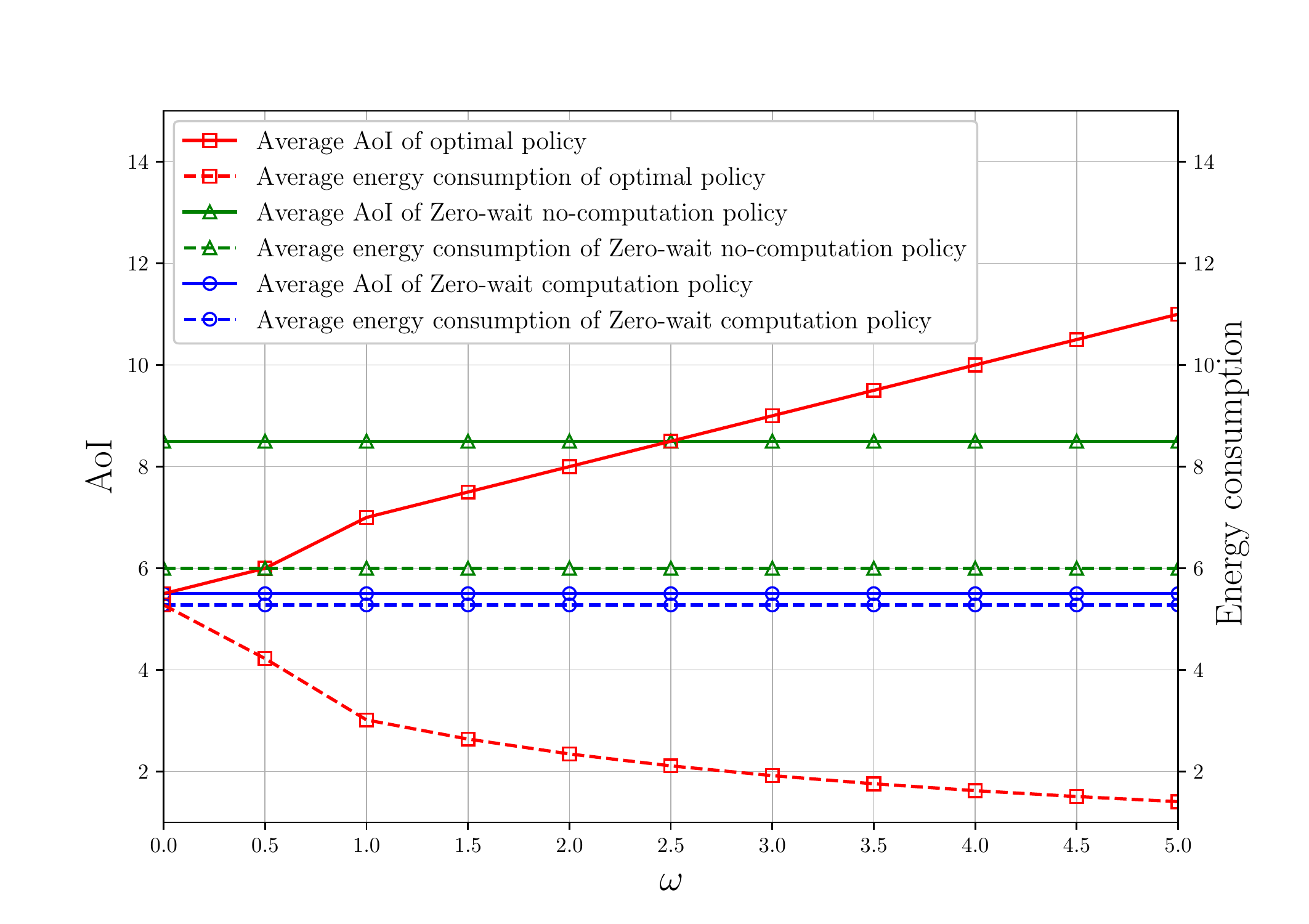}}\caption{\label{fig:opt-vers-zerowait-sp2-weight}Performance comparison among
the optimal policy, the zero-wait no-computation policy, and the zero-wait
computation policy ($T_{u}=6$, $T_{u}'=2$, $l=3$, $v=5$, $f=45$,
$\tau=1$, $\kappa=0.00005$, $P=6$). (a) The average cost versus
$\text{\ensuremath{\omega}}$. (b) The average AoI and the average
energy consumption versus $\omega$.}
\end{figure}

In Fig. \ref{fig:opt-vers-zerowait-sp2-weight}, the average cost
of the optimal policy and the two baseline policies are compared
with respect to $\omega$ when $T_{u}\geq T_{p}+T_{u}'$ and $C_{u}\geq\frac{T_{p}C_{p}+T_{u}'C_{u}}{T_{p}+T_{u}'}$.
As we can see from Fig. \ref{fig:opt-vers-zerowait-sp2-weight}(a),
the optimal policy outperforms both two baseline policies.  As the
weight factor increases, the average cost of the optimal policy and
the baseline policies increase but with different rates. From Fig.
\ref{fig:opt-vers-zerowait-sp2-weight}(b) we can see that the average
AoI and the energy consumption of the two zero-wait baseline policies
are constant for any $\omega$. The gap between the optimal policy
and the zero-wait baseline policies grows with $\omega$ in \ref{fig:opt-vers-zerowait-sp2-weight}(a),
because the zero-wait policies suffer from a higher weighted energy
consumption when the weighting factor is large. For the optimal policy,
we can see that, with the increase of $\omega$, its average energy
consumption tends to be smaller while the average AoI tends to be
larger. This is the result of the balance between the AoI reduction
and the energy consumption. Theorem \ref{lem:threshold-value} shows
that the threshold of the optimal policy in this case will increase
with $\omega$. Therefore, when $\omega$ is large, the IoT device
will update only when the AoI is large enough. 

\section{Conclusion}

In this paper, we have studied the problem for optimizing information
freshness in computing-enabled IoT systems by jointly controlling
the preprocessing and transmission at the IoT device. To minimize
the weighted sum of the average AoI associated with the destination
and the energy consumed by the IoT device, we have formulated an infinite
horizon average cost SMDP. By transforming the SMDP to an equivalent
uniform time step MDP, we have investigated the structure of the
optimal update policy and provided a structure-aware relative policy
iteration algorithm. We have further proved the switch-type structure
of the optimal policy in a special scenario where the status updates
are transmitted over a reliable channel. Simulation results have shown
that the optimal update policy can adjust adaptively based on the
channel quality and the degree of computational intensity of the status
update. By comparing the optimal update policy with two other zero-wait
policies, it is shown that the optimal update policy achieves a good
balance between the AoI reduction and the energy consumption.

\appendix{}

\subsection{\label{subsec:Proof-of-Lem1}Proof of Lemma \ref{lem:lem1}}

According to the value iteration algorithm (VIA) \cite[Chapter 4.3]{dimitrip.bertsekasDynamicProgrammingOptimal2007},
we prove Lemma \ref{lem:lem1} by using mathematical induction. We
first denote by $V_{k}(s)$ and $Q_{k}(s,\bm{a})$ the state value
function and state-action value functions at iteration $k$, respectively.
In particular, $Q_{k}(s,\bm{a})$ is defined as: 
\begin{equation}
Q_{k}(s,\boldsymbol{a})\triangleq\bar{R}(s,\boldsymbol{a})+\sum\limits _{s'\in\mathcal{S}}\bar{p}(s'\mid s,\boldsymbol{a})V_{k}(s'),\forall s\in\mathcal{S},
\end{equation}
where $s'$ is given by (\ref{eq:Dynamic}). For each state $s$,
VIA calculates $V_{k+1}(s)$ according to
\begin{equation}
V_{k+1}(s)=\min\limits _{\boldsymbol{a}\in\mathcal{A}}Q_{k}(s,\boldsymbol{a}).\label{eq:valuefun-min}
\end{equation}
Under any initialization of $V_{0}(s)$, the generated sequence $\{V_{k}(s)\}$
converges to $V(s)$, i.e., 
\begin{equation}
\lim_{k\rightarrow\infty}V_{k}(s)=V(s),\forall s\in\mathcal{S},\label{eq:value-limit}
\end{equation}
where $V(s)$ satisfies the Bellman equation in (\ref{eq:bellman-equ}).
Therefore, we can prove the monotonicity of $V(s)$ by showing that
this property is also possessed by $V_{k}(s)$ for any $k$. Particularly,
we need to prove that for any $s_{1},s_{2}\in\mathcal{S}$, such that
$s_{1}\le s_{2}$,
\begin{equation}
V_{k}(s_{1})\le V_{k}(s_{2}),\quad k=0,1,\ldots\label{eq:monotonicity}
\end{equation}

We initialize $V_{0}(s)=0$ for all $s\in\mathcal{S}$ without sacrificing
generality. Then, we prove (\ref{eq:monotonicity}) by using mathematical
induction. First, since $V_{0}(s)=0$ for all $s\in\mathcal{S}$,
(\ref{eq:monotonicity}) holds for $k=0$. Next, we assume that (\ref{eq:monotonicity})
holds up till $k>0$ and inspect whether it holds for $k+1$. 

When $\boldsymbol{a}=(0,0)$, we have $Q_{k}(s_{1},(0,0))=s_{1}+(1-\epsilon)V_{k}(s_{1})+\epsilon V_{k}(s_{1}+1)$
and $Q_{k}(s_{2},(0,0))=s_{2}+(1-\epsilon)V_{k}(s_{2})+\epsilon V_{k}(s_{2}+1)$.
Since $s_{1}\le s_{2}$, $V_{k}(s_{1})\le V_{k}(s_{2})$ and $V_{k}(s_{1}+1)\le V_{k}(s_{2}+1)$,
we can easily see that $Q_{k}(s_{1},(0,0))\le Q_{k}(s_{2},(0,0))$.

When $\boldsymbol{a}=(0,1)$ or $(1,1)$, the state-action value functions
at iteration $k$ are given by
\begin{alignat}{1}
Q_{k}(s_{1},\boldsymbol{a}) & =s_{1}+\frac{1}{2}(L(\boldsymbol{a})-1)+\omega\frac{C(\boldsymbol{a})}{L(\boldsymbol{a})}\nonumber \\
 & +\left(1-\frac{\epsilon}{L(\boldsymbol{a})}\right)V_{k}(s_{1})+\frac{\epsilon}{L(\boldsymbol{a})}p_{s}^{L_{u}(\boldsymbol{a})}V_{k}(L(\boldsymbol{a}))\nonumber \\
 & +\frac{\epsilon}{L(\boldsymbol{a})}(1-p_{s}^{L_{u}(\boldsymbol{a})})V_{k}(s_{1}L(\boldsymbol{a}))
\end{alignat}
and
\begin{alignat}{1}
Q_{k}(s_{2},\boldsymbol{a}) & =s_{2}+\frac{1}{2}(L(\boldsymbol{a})-1)+\omega\frac{C(\boldsymbol{a})}{L(\boldsymbol{a})}\nonumber \\
 & +\left(1-\frac{\epsilon}{L(\boldsymbol{a})}\right)V_{k}(s_{2})+\frac{\epsilon}{L(\boldsymbol{a})}p_{s}^{L_{u}(\boldsymbol{a})}V_{k}(L(\boldsymbol{a}))\nonumber \\
 & +\frac{\epsilon}{L(\boldsymbol{a})}(1-p_{s}^{L_{u}(\boldsymbol{a})})V_{k}(s_{2}L(\boldsymbol{a})).
\end{alignat}
Keeping in mind that $V_{k}(s_{1})\le V_{k}(s_{2})$, we can prove
that $Q_{k}(s_{1},(0,1))\le Q_{k}(s_{2},(0,1))$ and $Q_{k}(s_{1},(1,1))\le Q_{k}(s_{2},(1,1))$. 

According to (\ref{eq:valuefun-min}), we can deduce that $V_{k+1}(s_{1})\text{\ensuremath{\le}}V_{k+1}(s_{2})$,
i.e., (\ref{eq:monotonicity}) holds for $k+1$. By induction, we
can show that (\ref{eq:monotonicity}) holds for any $k$. This concludes
our proof.

\subsection{\label{subsec:Proof-of-Lem2}Proof of Lemma \ref{lem:lem2}}

The proof follows the same procedure of Lemma \ref{lem:lem1}. The
concavity of $V(s)$ in $s$ can be proved by showing that for any
$s_{1},s_{2}\in\mathcal{S}$ and $w\in N$, such that $s_{1}\le s_{2}$,
\begin{equation}
V_{k}(s_{1}+w)-V_{k}(s_{1})\geq V_{k}(s_{2}+w)-V_{k}(s_{2}),\quad k=0,1,\ldots\label{eq:multimodularity}
\end{equation}

We initialize $V_{0}(s)=0$ for all $s\in\mathcal{S}$ without sacrificing
generality. Thus, (\ref{eq:multimodularity}) holds for $k=0$. Next,
we assume that (\ref{eq:multimodularity}) holds up till $k>0$ and
inspect whether it holds for $k+1$. For convenience, we now define
$\Delta Q(s,s',\boldsymbol{a})=Q(s,\boldsymbol{a})-Q(s',\boldsymbol{a})$. 

When $\boldsymbol{a}=(0,0)$, we have
\begin{flalign}
 & \Delta Q_{k}(s_{1}+w,s_{1},(0,0))-\Delta Q_{k}(s_{2}+w,s_{2},(0,0))\nonumber \\
= & (1-\epsilon)\left(\left(V_{k}(s_{1}+w)-V_{k}(s_{1})\right)-\left(V_{k}(s_{2}+w)-V_{k}(s_{2})\right)\right)\nonumber \\
 & +\epsilon(\left(V_{k}(s_{1}+w+1)-V_{k}(s_{1}+1)\right)-(V_{k}(s_{2}+w+1)\nonumber \\
 & -V_{k}(s_{2}+1))).
\end{flalign}
Since $V_{k}(s_{1}+w)-V_{k}(s_{1})\ge V_{k}(s_{2}+w)-V_{k}(s_{2})$
and $V_{k}(s_{1}+w+1)-V_{k}(s_{1}+1)\ge V_{k}(s_{2}+w+1)-V_{k}(s_{2}+1)$,
we can easily see that $\Delta Q_{k}(s_{1}+w,s_{1},(0,0))-\Delta Q_{k}(s_{2}+w,s_{2},(0,0))\ge0$.
Hence, $Q_{k}(s,(0,0))$ is concave in $s$. 

Similarly, for actions $(0,1)$ and $(1,1)$, we have
\begin{flalign}
 & \Delta Q_{k}(s_{1}+w,s_{1},(0,1))-\Delta Q_{k}(s_{2}+w,s_{2},(0,1))\nonumber \\
= & \left(1-\frac{\epsilon}{T_{u}}\right)((V_{k}(s_{1}+w)-V_{k}(s_{1}))-(V_{k}(s_{2}+w)\nonumber \\
- & V_{k}(s_{2})))+\frac{\epsilon}{T_{u}}(1-p_{s}^{T_{u}})((V_{k}(s_{1}+w+T_{u})\nonumber \\
- & V_{k}(s_{1}+T_{u}))-(V_{k}(s_{2}+w+T_{u})-V_{k}(s_{2}+T_{u}))),
\end{flalign}
and
\begin{flalign}
 & \Delta Q_{k}(s_{1}+w,s_{1},(1,1))-\Delta Q_{k}(s_{2}+w,s_{2},(1,1))\nonumber \\
= & \left(1-\frac{\epsilon}{T_{p}+T_{u}'}\right)((V_{k}(s_{1}+w)-V_{k}(s_{1}))-(V_{k}(s_{2}+w)\nonumber \\
- & V_{k}(s_{2})))+\frac{\epsilon}{T_{p}+T_{u}'}(1-p_{s}^{T_{u}'})((V_{k}(s_{1}+w+T_{p}+T_{u}')\nonumber \\
- & V_{k}(s_{1}+T_{p}+T_{u}'))-(V_{k}(s_{2}+w+T_{p}+T_{u}')\nonumber \\
- & V_{k}(s_{2}+T_{p}+T_{u}'))).
\end{flalign}
Since $V_{k}(s_{1}+w)-V_{k}(s_{1})\ge V_{k}(s_{2}+w)-V_{k}(s_{2})$,
$V_{k}(s_{1}+w+T_{u})-V_{k}(s_{1}+T_{u})\ge V_{k}(s_{2}+w+T_{u})-V_{k}(s_{2}+T_{u})$,
and $V_{k}(s_{1}+w+T_{p}+T_{u}')-V_{k}(s_{1}+T_{p}+T_{u}')\ge V_{k}(s_{2}+w+T_{p}+T_{u}')-V_{k}(s_{2}+T_{p}+T_{u}')$,
we can also verify that $\Delta Q_{k}(s_{1}+w,s_{1},(0,1))-\Delta Q_{k}(s_{2}+w,s_{2},(0,1))\ge0$
and $\Delta Q_{k}(s_{1}+w,s_{1},(1,1))-\Delta Q_{k}(s_{2}+w,s_{2},(1,1))\ge0$.
Therefore, both $Q_{k}(s,(0,1))$ and $Q_{k}(s,(1,1))$ are also concave
in $s$. 

Since the value function $V_{k+1}(s)$ is the minimum of three concave
functions, it is also concave in $s$. Thus, we have $V_{k+1}(s_{1}+w)-V_{k+1}(s_{1})\ge V_{k+1}(s_{2}+w)-V_{k+1}(s_{2})$,
i.e., (\ref{eq:multimodularity}) holds for $k+1$. Therefore, by
induction, we can show that (\ref{eq:multimodularity}) holds for
any $k$. This concludes our proof.

\subsection{\label{subsec:Proof-of-Lem3}Proof of Lemma \ref{lem:lem3}}

The proof follows the same procedure of Lemma \ref{lem:lem1}. The
lower bound of $V(s_{2})-V(s_{1})$ can be proved by showing that
for any $s_{1},s_{2}\in\mathcal{S}$, such that $s_{1}\le s_{2}$,
\begin{equation}
V_{k}(s_{2})-V_{k}(s_{1})\geq\frac{L(\boldsymbol{a}_{f})}{\epsilon p_{s}^{L_{u}(\boldsymbol{a}_{f})}}(s_{2}-s_{1}),\quad k=0,1,\ldots\label{eq:slopebound}
\end{equation}

We initialize $V_{0}(s)=\frac{L(\boldsymbol{a}_{f})}{\epsilon p_{s}^{L_{u}(\boldsymbol{a}_{f})}}s$
for all $s\in\mathcal{S}$ without sacrificing generality. Thus,
(\ref{eq:slopebound}) holds for $k=0$. Next, we assume that (\ref{eq:slopebound})
holds up till $k>0$ and hence we have $V_{k}(s_{2})-V_{k}(s_{1})\geq\frac{L(\boldsymbol{a}_{f})}{\epsilon p_{s}^{L_{u}(\boldsymbol{a}_{f})}}(s_{2}-s_{1})$
and $V_{k}(s_{2}+1)-V_{k}(s_{1}+1)\geq\frac{L(\boldsymbol{a}_{f})}{\epsilon p_{s}^{L_{u}(\boldsymbol{a}_{f})}}(s_{2}-s_{1})$. 

Then, we inspect whether it holds for $k+1$. We first consider the
case when $\boldsymbol{a}_{f}=(0,1)$ and we have $\frac{L(\boldsymbol{a}_{f})}{\epsilon p_{s}^{L_{u}(\boldsymbol{a}_{f})}}=\frac{T_{u}}{\epsilon p_{s}^{T_{u}}}$.
Since $V_{k+1}(s)=\min\limits _{\boldsymbol{a}\in\mathcal{A}}Q_{k}(s,\boldsymbol{a})$,
we investigate the three state-action value functions, in the following,
respectively. 

When $\boldsymbol{a}=(0,0)$, we have
\begin{flalign}
 & \Delta Q_{k}(s_{2},s_{1},(0,0))\nonumber \\
 & =(s_{2}-s_{1})+(1-\epsilon)\left(V_{k}(s_{2})-V_{k}(s_{1})\right)\nonumber \\
 & \quad+\epsilon\left(V_{k}(s_{2}+1)-V_{k}(s_{1}+1)\right)\nonumber \\
 & \geq(s_{2}-s_{1})+\frac{L(\boldsymbol{a}_{f})}{\epsilon p_{s}^{L_{u}(\boldsymbol{a}_{f})}}(s_{2}-s_{1})\nonumber \\
 & =\left(1+\frac{L(\boldsymbol{a}_{f})}{\epsilon p_{s}^{L_{u}(\boldsymbol{a}_{f})}}\right)(s_{2}-s_{1})\geq\frac{L(\boldsymbol{a}_{f})}{\epsilon p_{s}^{L_{u}(\boldsymbol{a}_{f})}}(s_{2}-s_{1}).
\end{flalign}
When $\boldsymbol{a}=(0,1)$, we have
\begin{flalign}
 & \Delta Q_{k}(s_{2},s_{1},(0,1))\nonumber \\
 & =(s_{2}-s_{1})+\left(1-\frac{\epsilon}{T_{u}}\right)\left(V_{k}(s_{2})-V_{k}(s_{1})\right)\nonumber \\
 & \text{\ensuremath{\quad}}+\frac{\epsilon}{T_{u}}\left(1-p_{s}^{T_{u}}\right)\left(V_{k}(s_{2}+T_{u})-V_{k}(s_{1}+T_{u})\right)\nonumber \\
 & \geq(s_{2}-s_{1})+\left(1-\frac{\epsilon}{T_{u}}p_{s}^{T_{u}}\right)\frac{T_{u}}{\epsilon p_{s}^{T_{u}}}(s_{2}-s_{1})\nonumber \\
 & =\frac{T_{u}}{\epsilon p_{s}^{T_{u}}}(s_{2}-s_{1}),
\end{flalign}
When $\boldsymbol{a}=(1,1)$, we have
\begin{flalign}
 & \Delta Q_{k}(s_{2},s_{1},(1,1))\nonumber \\
 & =(s_{2}-s_{1})+\left(1-\frac{\epsilon}{T_{p}+T_{u}'}\right)\left(V_{k}(s_{2})-V_{k}(s_{1})\right)\nonumber \\
 & \quad+\frac{\epsilon}{T_{p}+T_{u}'}\left(1-p_{s}^{T_{u}'}\right)(V_{k}(s_{2}+T_{p}+T_{u}')\nonumber \\
 & \quad-V_{k}(s_{1}+T_{p}+T_{u}'))\nonumber \\
 & \geq(s_{2}-s_{1})+\left(1-\frac{\epsilon}{T_{p}+T_{u}'}p_{s}^{T_{u}'}\right)\frac{T_{u}}{\epsilon p_{s}^{T_{u}}}(s_{2}-s_{1})\nonumber \\
 & =\frac{T_{u}}{\epsilon p_{s}^{T_{u}}}(s_{2}-s_{1})+\left(1-\frac{T_{u}}{T_{p}+T_{u}'}p_{s}^{T_{u}'-T_{u}}\right)(s_{2}-s_{1})\nonumber \\
 & \geq\frac{T_{u}}{\epsilon p_{s}^{T_{u}}}(s_{2}-s_{1}).
\end{flalign}
Therefore, we can prove that $V_{k}(s_{2})-V_{k}(s_{1})\geq\frac{T_{u}}{\epsilon p_{s}^{T_{u}}}(s_{2}-s_{1})$
for any $k$ when the optimal actions in $s_{1}$ and $s_{2}$ are
the same. 

When the optimal policy in $s_{1}$ and $s_{2}$ are two different
actions, i.e., $\boldsymbol{a}_{1}$ and $\boldsymbol{a}_{2}$, we
have
\begin{align}
V_{k}(s_{2})-V_{k}(s_{1})= & Q_{k}(s_{2},a_{2})-Q_{k}(s_{1},a_{1})\nonumber \\
\geq & Q_{k}(s_{2},a_{2})-Q_{k}(s_{1},a_{2})\nonumber \\
\geq & \frac{T_{u}}{\epsilon p_{s}^{T_{u}}}(s_{2}-s_{1}).
\end{align}
Therefore, we can also verify that $V_{k}(s_{2})-V_{k}(s_{1})\geq\frac{T_{u}}{\epsilon p_{s}^{T_{u}}}(s_{2}-s_{1})$
for any $k$ in this case.

Altogether, we can conclude that $V_{k}(s_{2})-V_{k}(s_{1})\geq\frac{T_{u}}{\epsilon p_{s}^{T_{u}}}(s_{2}-s_{1})$
for any $k$. By induction, we have $V(s_{2})-V(s_{1})\geq\frac{T_{u}}{\epsilon}(s_{2}-s_{1})$.
By following the same analysis as the one done when $\boldsymbol{a}_{f}=(0,1)$,
we can prove that $V_{k}(s_{2})-V_{k}(s_{1})\geq\frac{T_{p}+T_{u}'}{\epsilon p_{s}^{T_{u}'}}(s_{2}-s_{1})$
when $\boldsymbol{a}_{f}=(1,1)$. This concludes our proof.

\subsection{\label{subsec:Proof-of-thm4}Proof of Theorem \ref{thm:threshold-structure}}

To proceed with the proof, we provide the following lemma that will
be useful to our proof. 
\begin{lem}
\label{lem:lem-n}For any $s_{2},s_{1}\in\mathcal{S}$, such that
$s_{2}\geq s_{1}$, $V(s_{2})-V(s_{1})\geq\Delta Q(s_{2},s_{1},\boldsymbol{a}_{f})$. 
\end{lem}
\begin{IEEEproof}
For any $s_{2},s_{1}\in\mathcal{S}$, such that $s_{2}\geq s_{1}$,
we have 
\begin{flalign}
 & \Delta Q(s_{2},s_{1},\boldsymbol{a}_{f})-\left(V(s_{2})-V(s_{1})\right)\nonumber \\
= & (s_{2}-s_{1})+\frac{\epsilon}{L(\boldsymbol{a}_{f})}\left(1-p_{s}^{L_{u}(\boldsymbol{a}_{f})}\right)(V(s_{2}+L(\boldsymbol{a}_{f}))\nonumber \\
 & -V(s_{1}+L(\boldsymbol{a}_{f})))-\frac{\epsilon}{L(\boldsymbol{a}_{f})}(V(s_{2})-V(s_{1})).
\end{flalign}
Since the concavity of $V(s)$ has been proved in Lemma \ref{lem:lem2},
we can easily see that $V(s_{2}+L(\boldsymbol{a}_{f}))-V(s_{1}+L(\boldsymbol{a}_{f}))\leq V(s_{2})-V(s_{1})$.
Therefore, we have
\begin{flalign}
 & \Delta Q(s_{2},s_{1},\boldsymbol{a}_{f})-\left(V(s_{2})-V(s_{1})\right)\nonumber \\
\leq & (s_{2}-s_{1})+\frac{\epsilon}{L(\boldsymbol{a}_{f})}\left(1-p_{s}^{L_{u}(\boldsymbol{a}_{f})}\right)(V(s_{2})-V(s_{1}))\nonumber \\
 & -\frac{\epsilon}{L(\boldsymbol{a}_{f})}(V(s_{2})-V(s_{1})).\nonumber \\
= & (s_{2}-s_{1})-\frac{\epsilon p_{s}^{L_{u}(\boldsymbol{a}_{f})}}{L(\boldsymbol{a}_{f})}(V(s_{2})-V(s_{1})).\label{eq:submodularity}
\end{flalign}
As proved in Lemma \ref{lem:lem3} that $V(s_{2})-V(s_{1})\geq\frac{L(\boldsymbol{a}_{f})}{\epsilon p_{s}^{L_{u}(\boldsymbol{a}_{f})}}(s_{2}-s_{1})$,
it is easy to see that $\Delta Q(s_{2},s_{1},\boldsymbol{a}_{f})-\left(V(s_{2})-V(s_{1})\right)\leq0$.
This completes the proof of Lemma \ref{lem:lem-n}. 
\end{IEEEproof}
Now we can prove the threshold structure of the optimal policy. Suppose
$s_{2}\geq s_{1}$ and $\pi^{*}(s_{1})=\boldsymbol{a}_{f}$, it is
easily to see that $V(s_{1})=Q(s_{1},\boldsymbol{a}_{f})$. According
to Lemma \ref{lem:lem-n}, we know that $V(s_{2})-V(s_{1})\geq Q(s_{2},\boldsymbol{a}_{f})-Q(s_{1},\boldsymbol{a}_{f})$.
Therefore, we have $V(s_{2})\geq Q(s_{2},\boldsymbol{a}_{f})$. Since
the value function is a minimum of three state-action cost functions,
we have $V(s_{2})\leq Q(s_{2},\boldsymbol{a}_{f})$. Altogether, we
can assert that $V(s_{2})=Q(s_{2},\boldsymbol{a}_{f})$ and $\pi^{*}(s_{2})=\boldsymbol{a}_{f}$. 

\subsection{\label{subsec:Proof-of-lem:simplify1}Proof of Lemma \ref{lem:simplify1}}

We first consider the case that $\boldsymbol{a}_{f}=(1,1)$. In this
case, we have $L(\bm{a}_{f})=T_{p}+T_{u}'$, $C(\bm{a}_{f})=T_{p}C_{p}+T_{u}'C_{u}$,
and $\mathcal{S}^{\dagger}\triangleq\left\{ T_{p}+T_{u}',\cdots,\hat{\delta}\right\} $.
According to Lemma \ref{lem:lem3}, the state-action value function
with action $(0,0)$ can be expressed as 
\begin{align}
 & Q(s,(0,0))\nonumber \\
= & s+(1-\epsilon)V(s)+\epsilon V(s+1)\nonumber \\
\ge & s+(1-\epsilon)V(s)+\epsilon\left(V(s)+\frac{T_{p}+T_{u}'}{\epsilon}\right)\nonumber \\
= & s+T_{p}+T_{u}'+V(s).
\end{align}
In accordance with the definition of $Q(s,\boldsymbol{a})$, we have
\begin{align}
 & Q(s,(0,0))-Q(s,(1,1))\nonumber \\
\ge & \left(s+T_{p}+T_{u}'+V(s)\right)-\bigg(s+\frac{1}{2}(T_{p}+T_{u}'-1)\nonumber \\
 & +\omega\frac{T_{p}C_{p}+T_{u}'C_{u}}{T_{p}+T_{u}'}+\left(1-\frac{\epsilon}{T_{p}+T_{u}'}\right)V(s)\nonumber \\
 & +\frac{\epsilon}{T_{p}+T_{u}'}V(T_{p}+T_{u}')\bigg)\nonumber \\
= & \frac{1}{2}(T_{p}+T_{u}'+1)-\omega\frac{T_{p}C_{p}+T_{u}'C_{u}}{T_{p}+T_{u}'}\nonumber \\
 & +\frac{\epsilon}{T_{p}+T_{u}'}(V(s)-V(T_{p}+T_{u}')).
\end{align}
From Lemma \ref{lem:lem1}, we know that $V(s)\ge V(T_{p}+T_{u}')$
for $s\in\mathcal{S}^{\dagger}$. It is easy to see that $Q(s,(0,0))\geq Q(s,(1,1))$
for any $s\in\mathcal{S}^{\dagger}$ when $\frac{1}{2}(T_{p}+T_{u}'+1)\geq\omega\frac{T_{p}C_{p}+T_{u}'C_{u}}{T_{p}+T_{u}'}$.
Therefore, action $(1,1)$ is always better than action $(0,0)$. 

Similarly, we can also prove that $Q(s,(0,0))\geq Q(s,(0,1))$ for
any $s\in\mathcal{S}^{\dagger}$ when $\boldsymbol{a}_{f}=(0,1)$
and $\frac{1}{2}(T_{u}+1)\geq\omega C_{u}$, i.e., $(0,1)$ is always
better than $(0,0)$. 

\subsection{\label{subsec:Proof-of-threshold-range}Proof of Lemma \ref{lem:threshold-range}}

According to the Markov chains in Fig. \ref{fig:MarkovChain-1} and
the definition of the average cost in (\ref{eq:Problem}), the average
cost for $\Omega=L(\boldsymbol{a}_{1})$ is given by
\begin{align}
J_{1}= & \frac{R(L(\boldsymbol{a}_{1}),\boldsymbol{a}_{1})}{L(\boldsymbol{a}_{1})}\nonumber \\
= & \frac{3}{2}L(\boldsymbol{a}_{1})+\omega\frac{C(\boldsymbol{a}_{1})}{L(\boldsymbol{a}_{1})}-\frac{1}{2}.
\end{align}
Similarly, the average cost for $L(\boldsymbol{a}_{1})<\Omega\leq L(\boldsymbol{a}_{2})$
is given by
\begin{align}
J_{2}= & \frac{R(L(\boldsymbol{a}_{1}),\boldsymbol{a}_{2})+R(L(\boldsymbol{a}_{2}),\boldsymbol{a}_{1})}{L(\boldsymbol{a}_{1})+L(\boldsymbol{a}_{2})}\nonumber \\
= & L(\boldsymbol{a}_{1})L(\boldsymbol{a}_{2})+\omega\frac{C(\boldsymbol{a}_{1})+C(\boldsymbol{a}_{2})}{L(\boldsymbol{a}_{1})+L(\boldsymbol{a}_{2})},
\end{align}
and the average cost for $\Omega>L(\boldsymbol{a}_{2})$ is given
by
\begin{align}
J_{3}= & \frac{R(L(\boldsymbol{a}_{2}),\boldsymbol{a}_{2})}{L(\boldsymbol{a}_{2})}\nonumber \\
= & \frac{3}{2}L(\boldsymbol{a}_{2})+\omega\frac{C(\boldsymbol{a}_{2})}{L(\boldsymbol{a}_{2})}-\frac{1}{2}.
\end{align}
This concludes our proof.

\subsection{\label{subsec:Proof-of-simplify2}Proof of Lemma \ref{lem:simplify2}}

We first consider the case that $T_{u}\ge T_{p}+T_{u}'$ and $C_{u}\ge\frac{T_{p}C_{p}+T_{u}'C_{u}}{T_{p}+T_{u}'}$.
When $s=T_{p}+T_{u}'$, we have
\begin{flalign}
 & Q(T_{p}+T_{u}',(0,1))-Q(T_{p}+T_{u}',(1,1))\nonumber \\
= & \frac{1}{2}\left(T_{u}-(T_{p}+T_{u}')\right)+\omega\left(C_{u}-\frac{T_{p}C_{p}+T_{u}'C_{u}}{T_{p}+T_{u}'}\right)\nonumber \\
 & +\left(\frac{\epsilon}{T_{p}+T_{u}'}-\frac{\epsilon}{T_{u}}\right)V(T_{p}+T_{u}')\nonumber \\
 & +\left(\frac{\epsilon}{T_{u}}V(T_{u})-\frac{\epsilon}{T_{p}+T_{u}'}V(T_{p}+T_{u}')\right)\nonumber \\
\geq & \frac{\epsilon}{T_{u}}\left(V(T_{u})-V(T_{p}+T_{u}')\right).
\end{flalign}
From Lemma \ref{lem:lem1} we know that $V(T_{u})\ge V(T_{p}+T_{u}')$.
Therefore, we have $Q(T_{p}+T_{u}',(0,1))\ge Q(T_{p}+T_{u}',(1,1))$. 

Then, we will prove that $Q(s,\boldsymbol{a})$ is of a sub-modular
structure for $(0,1)$ and $(1,1)$, that is
\begin{equation}
Q(s_{1},(0,1))-Q(s_{1},(1,1))\leq Q(s_{2},(0,1))-Q(s_{2},(1,1)),\label{eq:sub-modular}
\end{equation}
for any $s_{2},s_{1}\in\mathcal{S}^{\dagger}$, such that $s_{2}\geq s_{1}$.
According to the definition of $Q(s,\boldsymbol{a})$, we have
\begin{align}
 & \left(Q(s_{1},(0,1))-Q(s_{1},(1,1))\right)\nonumber \\
 & -\left(Q(s_{2},(0,1))-Q(s_{2},(1,1))\right)\nonumber \\
= & \left((s_{1}-s_{2})+\left(1-\frac{\epsilon}{T_{u}}\right)(V(s_{1})-V(s_{2}))\right)\nonumber \\
 & -\left((s_{1}-s_{2})+\left(1-\frac{\epsilon}{T_{p}+T_{u}'}\right)(V(s_{1})-V(s_{2}))\right)\nonumber \\
= & \left(\frac{\epsilon}{T_{p}+T_{u}'}-\frac{\epsilon}{T_{u}}\right)(V(s_{1})-V(s_{2})).
\end{align}
Since $V(s_{1})\leq V(s_{2})$, we can easily see that (\ref{eq:sub-modular})
holds. Since $Q(T_{p}+T_{u}',(0,1))\ge Q(T_{p}+T_{u}',(1,1))$, we
can see that $Q(s,(0,1))\geq Q(s,(1,1))$ holds for any $s\in\mathcal{S}^{\dagger}$
when $T_{u}\geq T_{p}+T_{u}'$ and $C_{u}\geq\frac{T_{p}C_{p}+T_{u}'C_{u}}{T_{p}+T_{u}'}$,
i.e., $(1,1)$ is always better than $(0,1)$. 

Similarly, we can also prove that $Q(s,(0,1))\leq Q(s,(1,1))$ for
any $s\in\mathcal{S}^{\dagger}$ when $T_{u}\leq T_{p}+T_{u}'$ and
$C_{u}\leq\frac{T_{p}C_{p}+T_{u}'C_{u}}{T_{p}+T_{u}'}$, i.e., $(0,1)$
is always better than $(1,1)$. 

\subsection{\label{subsec:Proof-of-threshold-value}Proof of Lemma \ref{lem:threshold-value}}

\begin{figure}[tp]
\centering

\includegraphics[width=0.5\textwidth]{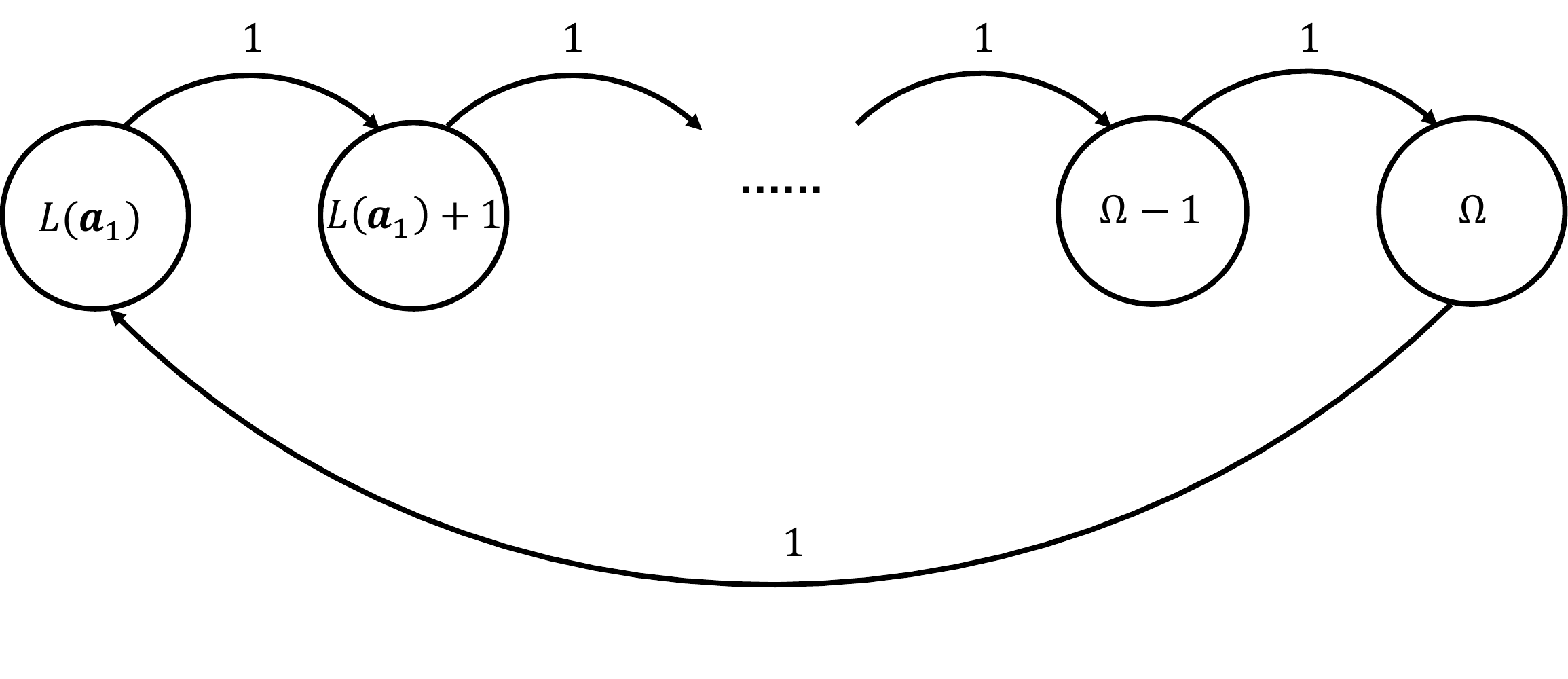}\caption{\label{fig:MarkovChain}The states transitions under a threshold policy
with the threshold of $\Omega$.}
\end{figure}

The MDP can be modeled via a DTMC with the same states for any threshold
policy of the type in Theorem \ref{thm:threshold2}, which is illustrated
in Fig. \ref{fig:MarkovChain}. We can easily see from Fig. \ref{fig:MarkovChain}
that this Markov chain is of a cyclic structure, in which the state
will cycle back and forth in the states between $L(\boldsymbol{a}_{1})$
and $\Omega$. 

In each cycle, the device will take $\Omega-L(\boldsymbol{a}_{1})$
minislots to stay idle in the states from $L(\boldsymbol{a}_{1})$
to $\Omega-1$. Once the AoI reaches the threshold, a new status update
will be generated and action $\bm{a}_{1}$ is taken, which takes $L(\boldsymbol{a}_{1})$
minislots. Therefore, each cycle takes $\Omega$ minislots in total.
With the cyclic structure, we can obtain the average cost by finding
the average cost in a cycle. 

The average cost resulted by keeping idle in a cycle is given by
\begin{equation}
J_{1}(\Omega)=\frac{1}{\Omega}\stackrel[i=L(\boldsymbol{a}_{1})]{\Omega-1}{\sum}i.
\end{equation}
The average cost resulted by updating in a cycle is given by
\begin{equation}
J_{2}(\Omega)=\frac{1}{\Omega}R(\Omega,\boldsymbol{a}_{1}).
\end{equation}
Then, the average cost obtained by the threshold policy in Theorem
\ref{thm:threshold2} can be given by
\begin{align}
J(\Omega)= & J_{1}(\Omega)+J_{2}(\Omega)\nonumber \\
= & \frac{1}{\Omega}\left(\stackrel[i=L(\boldsymbol{a}_{1})]{\Omega-1}{\sum}i+R(\Omega,\boldsymbol{a}_{1})\right)\nonumber \\
= & \frac{1}{\Omega}\left(\stackrel[i=L(\boldsymbol{a}_{1})]{\Omega-1}{\sum}i+\stackrel[i=\Omega]{\Omega+L(\boldsymbol{a}_{1})-1}{\sum}i+\omega C(\boldsymbol{a}_{1})\right)\nonumber \\
= & \frac{1}{\Omega}\left(\stackrel[i=L(\boldsymbol{a}_{1})]{\Omega+L(\boldsymbol{a}_{1})-1}{\sum}i+\omega C(\boldsymbol{a}_{1})\right)\nonumber \\
= & L(\boldsymbol{a}_{1})+\frac{1}{2}(\Omega-1)+\frac{\omega C(\boldsymbol{a}_{1})}{\Omega}.
\end{align}

\subsection{\label{subsec:Proof-of-opt-threshold}Proof of Theorem \ref{thm:opt-threshold2}}

We derive the optimal threshold $\Omega^{*}$ by relaxing $\Omega$
to a continuous variable. We first calculate the second order derivative
of $J(\Omega)$ as follow, 
\begin{equation}
\frac{\partial^{2}J(\Omega)}{\partial\Omega^{2}}=\frac{2\omega C(\boldsymbol{a}_{1})}{\Omega^{3}}.
\end{equation}
It is easy to see that $\frac{\partial^{2}J(\Omega)}{\partial\Omega^{2}}\ge0$.
Therefore, $J(\Omega)$ is a convex function with respect to $\Omega$.
Then, we calculate the first order derivative of $J(\Omega)$ as follow,
\begin{equation}
\frac{\partial J(\Omega)}{\partial\Omega}=\frac{1}{2}-\frac{\omega C(\boldsymbol{a}_{1})}{\Omega^{2}}.
\end{equation}
By setting $\frac{\partial J(\Omega)}{\partial\Omega}$ to zero, we
can obtain the optimal threshold The solution to $\frac{\partial J(\Omega)}{\partial\Omega}=0$
is 
\begin{equation}
\Omega'=\sqrt{2\omega C(\bm{a}_{1})}.
\end{equation}
Since $\Omega'$ may not be an integer, the optimal threshold can
be expressed as 
\begin{equation}
\Omega^{*}=\arg\min\left(J(\left\lfloor \Omega'\right\rfloor ),J(\left\lceil \Omega'\right\rceil )\right).
\end{equation}

\bibliographystyle{IEEEtran}
\bibliography{AoI}

\end{document}